\documentclass[]{spie}  

\usepackage{amsmath,amsfonts,amssymb}
\usepackage{graphicx}
\usepackage{url}
\usepackage{comment}
\usepackage[colorlinks=true, allcolors=blue]{hyperref}

\title{Calibration and performance of the readout system based on switched capacitor arrays for the Large-Sized Telescope of the Cherenkov Telescope Array}

\author[a]{Seiya~Nozaki}
\author[b]{Kyosuke~Awai}
\author[c]{Aya Bamba}
\author[d]{Juan Abel Barrio}
\author[e]{Maria~Isabel~Bernardos}
\author[f]{Oscar~Blanch}
\author[f]{Joan~Boix}
\author[g]{Franca~Cassol}
\author[h]{Yuki~Choushi}
\author[e]{Carlos~Delgado}
\author[e]{Carlos Diaz}
\author[i]{Nadia Fouque}
\author[e]{Lluis Freixas}
\author[j]{Pawel Gliwny}
\author[k]{Shunichi Gunji}
\author[b]{Daniela~Hadasch}
\author[g]{Dirk~Hoffmann}
\author[g]{Julien~Houles}
\author[b]{Yusuke~Inome}
\author[b]{Yuki Iwamura}
\author[f]{Léa Jouvin}
\author[l]{Hideaki~Katagiri}
\author[h]{Kiomei~Kawamura}
\author[f]{Daniel~Kerszberg}
\author[a]{Yusuke~Konno}
\author[a]{Hidetoshi~Kubo}
\author[m]{Junko~Kushida}
\author[b]{Yukiho~Kobayashi}
\author[n]{Ruben~L\'{o}pez-Coto}
\author[e]{Gustavo~Martinez}
\author[a]{Shu Masuda}
\author[b]{Daniel~Mazin}
\author[f]{Abelardo~Moralejo}
\author[f]{Elena~Moretti}
\author[o]{Tsutomu~Nagayoshi}
\author[k]{Takeshi~Nakamori}
\author[m]{Kyoshi~Nishijima}
\author[l]{Yuto~Nogami}
\author[f]{Leyre~Nogu\'{e}s}
\author[b]{Hideyuki~Ohoka}
\author[a]{Tomohiko Oka}
\author[b]{Nao~Okazaki}
\author[p, q]{Akira~Okumura}
\author[r]{Reiko~Orito}
\author[i]{Jean-Luc~Panazol}
\author[s]{Riccardo~Paoletti}
\author[f]{Cristobal Pio}
\author[b]{Miguel Polo}
\author[i]{Julie Prast}
\author[b]{Takayuki~Saito}
\author[b]{Shunsuke~Sakurai}
\author[j]{Julian~Sitarek}
\author[o]{Yuji~Sunada}
\author[l]{Megumi~Suzuki}
\author[b]{Mitsunari Takahashi}
\author[h]{Kenji~Tamura}
\author[t]{Manobu~Tanaka}
\author[d]{Luis~Angel~Tejedor}
\author[o]{Yukikatsu~Terada}
\author[b, u]{Masahiro~Teshima}
\author[h]{Yusuke~Tsukamoto}
\author[h]{Tokonatsu~Yamamoto}

\affil[a]{Division of Physics and Astronomy, Graduate School of Science, Kyoto University, Sakyo-ku, Kyoto,
606-8502, Japan}
\affil[b]{Institute for Cosmic Ray Research, University of Tokyo, 5-1-5, Kashiwa-no-ha, Kashiwa, Chiba
277-8582, Japan}
\affil[c]{Department of Physics, Graduate School of Science, University of Tokyo, 7-3-1 Hongo, Bunkyo-ku,
Tokyo 113-0033, Japan}
\affil[d]{EMFTEL department and IPARCOS, Universidad Complutense de Madrid, E-28040 Madrid, Spain}
\affil[e]{CIEMAT, Avda. Complutense 40, 28040 Madrid, Spain}
\affil[f]{Institut de Fisica d’Altes Energies (IFAE), The Barcelona Institute of Science and Technology, Campus UAB, 08193 Bellaterra (Barcelona), Spain}
\affil[g]{Aix Marseille Univ, CNRS/IN2P3, CPPM, Marseille, France, 163 Avenue de Luminy, 13288 Marseille cedex 09, France}
\affil[h]{Department of Physics, Konan University, Kobe, Hyogo, 658-8501, Japan}
\affil[i]{Laboratoire d’Annecy-le-Vieux de Physique des Particules, Universite de Savoie Mont-Blanc, ´CNRS/IN2P3, 9 Chemin de Bellevue - BP 110, 74941 Annecy-le-Vieux Cedex, France}
\affil[j]{Faculty of Physics and Applied Computer Science, University of L\'{o}d\'{z}, ul. Pomorska 149-153, 90-236 L\'{o}d\'{z}, Poland}
\affil[k]{Department of Physics, Yamagata University, Yamagata, Yamagata 990-8560, Japan}
\affil[l]{Faculty of Science, Ibaraki University, Mito, Ibaraki, 310-8512, Japan}
\affil[m]{Department of Physics, Tokai University, 4-1-1, Kita-Kaname, Hiratsuka, Kanagawa 259-1292, Japan}
\affil[n]{INFN Sezione di Padova, Via Marzolo 8, 35131, Padova, Italy}
\affil[o]{Graduate School of Science and Engineering, Saitama University, 255 Simo-Ohkubo, Sakura-ku, Saitama city, Saitama 338-8570, Japan}
\affil[p]{Institute for Space-Earth Environmental Research, Nagoya University, Chikusa-ku, Nagoya 464-8601, Japan}
\affil[q]{Kobayashi-Maskawa Institute (KMI) for the Origin of Particles and the Universe, Nagoya University, Chikusa-ku, Nagoya 464-8602, Japan}
\affil[r]{Graduate School of Technology, Industrial and Social Sciences, Tokushima University, Tokushima 770-8506, Japan}
\affil[s]{INFN Sezione di Pisa, Largo Pontecorvo 3, 56217 Pisa, Italy}
\affil[t]{Institute of Particle and Nuclear Studies, KEK (High Energy Accelerator Research Organization), 1-1 Oho, Tsukuba, 305-0801, Japan}
\affil[u]{Max-Planck-Institut f\"{u}r Physik, F\"{o}hringer Ring 6, 80805 M\"{u}nchen, Germany}

\authorinfo{Further author information: (Send correspondence to Seiya Nozaki)\\
Seiya Nozaki: E-mail: nozaki@cr.scphys.kyoto-u.ac.jp, Telephone: +81 (0)75 753 3869}

\begin{document} 
\maketitle

\begin{abstract}
 The Cherenkov Telescope Array (CTA) is the next-generation ground-based very-high-energy gamma-ray observatory. The Large-Sized Telescope (LST) of CTA is designed to detect gamma rays between 20~GeV and a few TeV with a 23-meter diameter mirror. We have developed the focal plane camera of the first LST, which has 1855 photomultiplier tubes (PMTs) and the readout system which samples a PMT waveform at GHz with switched capacitor arrays, Domino Ring Sampler ver4 (DRS4). To measure the precise pulse charge and arrival time of Cherenkov signals, we developed a method to calibrate the output voltage of DRS4 and the sampling time interval, as well as an analysis method to correct the spike noise of DRS4. Since the first LST was inaugurated in 2018, we have performed the commissioning tests and calibrated the camera. We characterised the camera in terms of the charge pedestal under various conditions of the night sky background, the charge resolution of each pixel, the charge uniformity of the whole camera, and the time resolutions with a test pulse and calibration laser.

\end{abstract}


\section{INTRODUCTION}
\label{sec:intro} 
The Cherenkov Telescope Array\cite{cta_hp} (CTA) is the next-generation ground-based observatory for very-high-energy gamma rays. The CTA consists of three types of telescopes with different mirror areas to cover a wide energy range (20~GeV--300~TeV) with an order of magnitude higher sensitivity than the predecessors. Among those telescopes, the Large-Sized Telescope (LST) is designed to detect low-energy gamma rays between 20~GeV and a few TeV with a 23~m diameter mirror. To make the most of such a large light collection area (about 400~m$^{2}$), the focal plane camera must detect as much reflected Cherenkov light as possible. We have developed each camera component to meet the CTA performance requirements for more than ten years and performed quality-control tests before installing the camera to the telescope\cite{Konno2016NIMPA, Sakurai2019ICRC}.

\begin{figure}
    \centering
    \includegraphics[height=5cm]{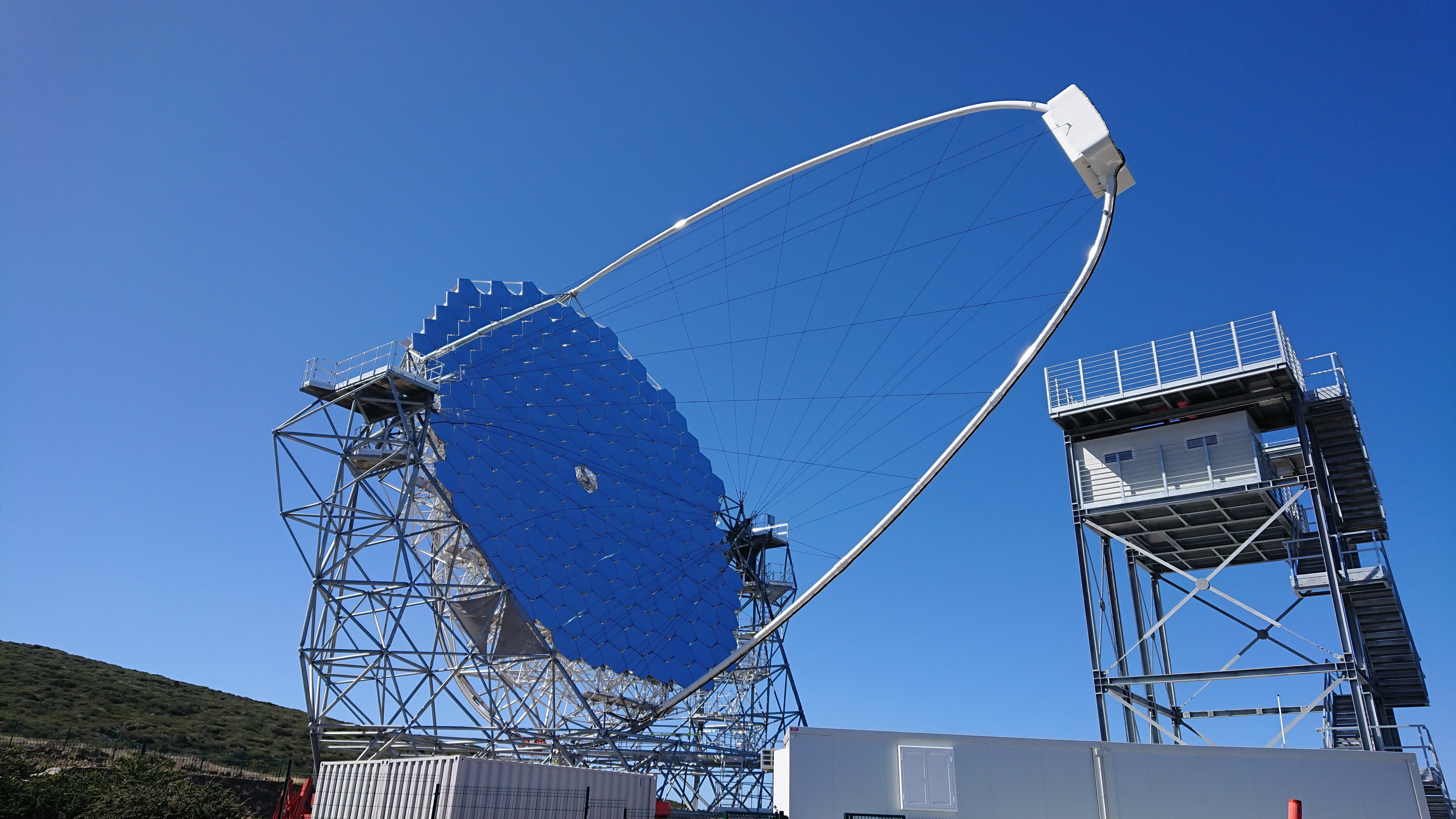}
    \caption{The first Large-Sized Telescope (LST-1).}
    \label{fig:LST_photo}
\end{figure}

The first LST (LST-1) was inaugurated in October 2018 in La Palma, Spain (Figure \ref{fig:LST_photo})\cite{Cortina2019ICRC_LST}. After the inauguration, various calibration tests were performed to adjust hardware parameters and verify the camera performance. In parallel, we have been developing the analysis software to extract physical parameters from low-level data, taking into account some intrinsic characteristics of the switched capacitor arrays, Domino Ring Sampler version 4 (DRS4), used for sampling the waveform of a Cherenkov signal. 
In this contribution, we describe the hardware design of the LST camera in Section \ref{sec:hardware_design}, a procedure for low-level calibration in Section \ref{sec:calibration}, and the readout performance of the LST camera after the hardware calibration with a dedicated analysis chain in Section \ref{sec:result}.

\section{Hardware Design of the LST camera}\label{sec:hardware_design}
\subsection{Overview}
\begin{figure}
  \begin{minipage}[b]{0.5\linewidth}
    \centering
    \includegraphics[clip, height=5.0cm]{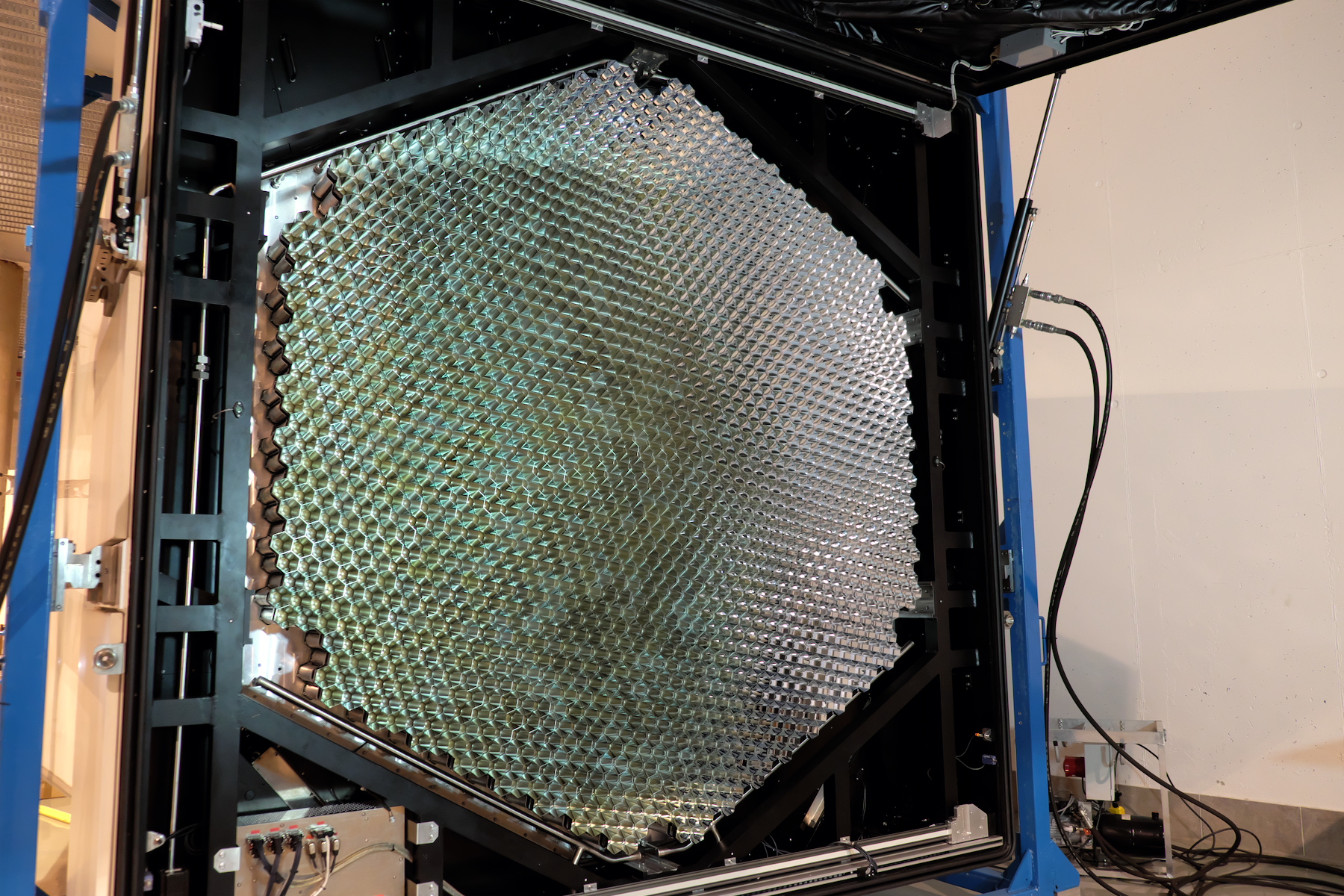}
  \end{minipage}
  \begin{minipage}[b]{0.5\linewidth}
    \centering
    \includegraphics[clip, height=5.0cm]{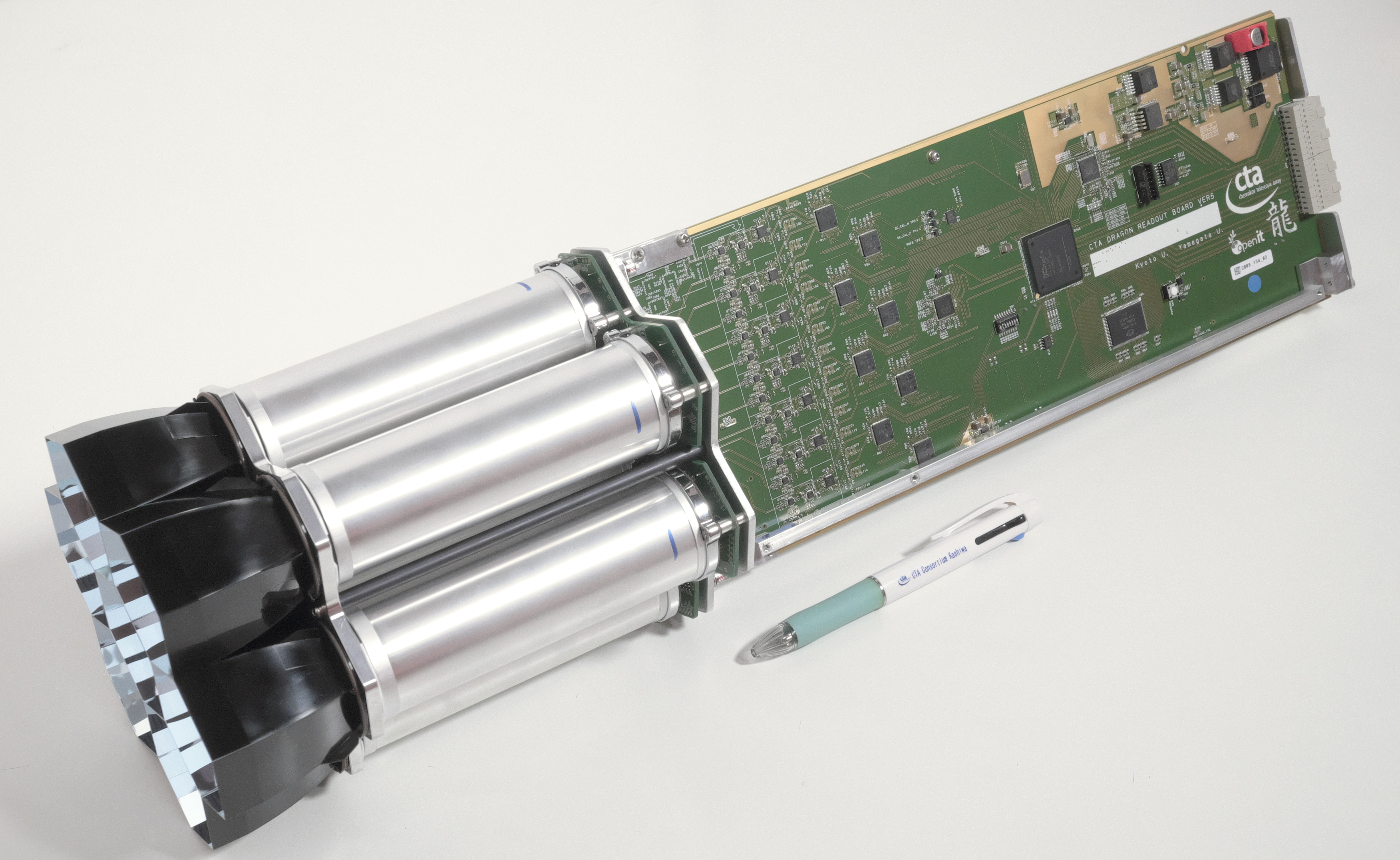}
  \end{minipage}
  \caption{The focal plane camera of the LST-1. Left: A frontal view of the camera. Right: A PMT module before installing in the LST-1 camera. Each PMT is equipped with a light guide, and seven PMTs are connected to a readout board based on the DRS4 chip. A trigger mezzanine is attached to the backside of the readout board. Backplane boards are located on the backside of the camera cabinet, and attached to the modules after module installation.}
  \label{fig:Camera_photo}
\end{figure}

Figure \ref{fig:Camera_photo} (left) shows a frontal view of the focal plane camera of the first LST. The camera has 1855 photomultiplier tubes (PMTs) organized in 265 modules, which are seven-pixel units shown in the Figure \ref{fig:Camera_photo} (right). Each module has seven PMTs with light guides, a slow control board, a readout board based on DRS4, a trigger mezzanine, and a backplane board. Besides, a single Trigger Interface Board (TIB) and a cooling system are in the LST camera structure.

\begin{figure}
    \centering
    \includegraphics[width=0.7\linewidth]{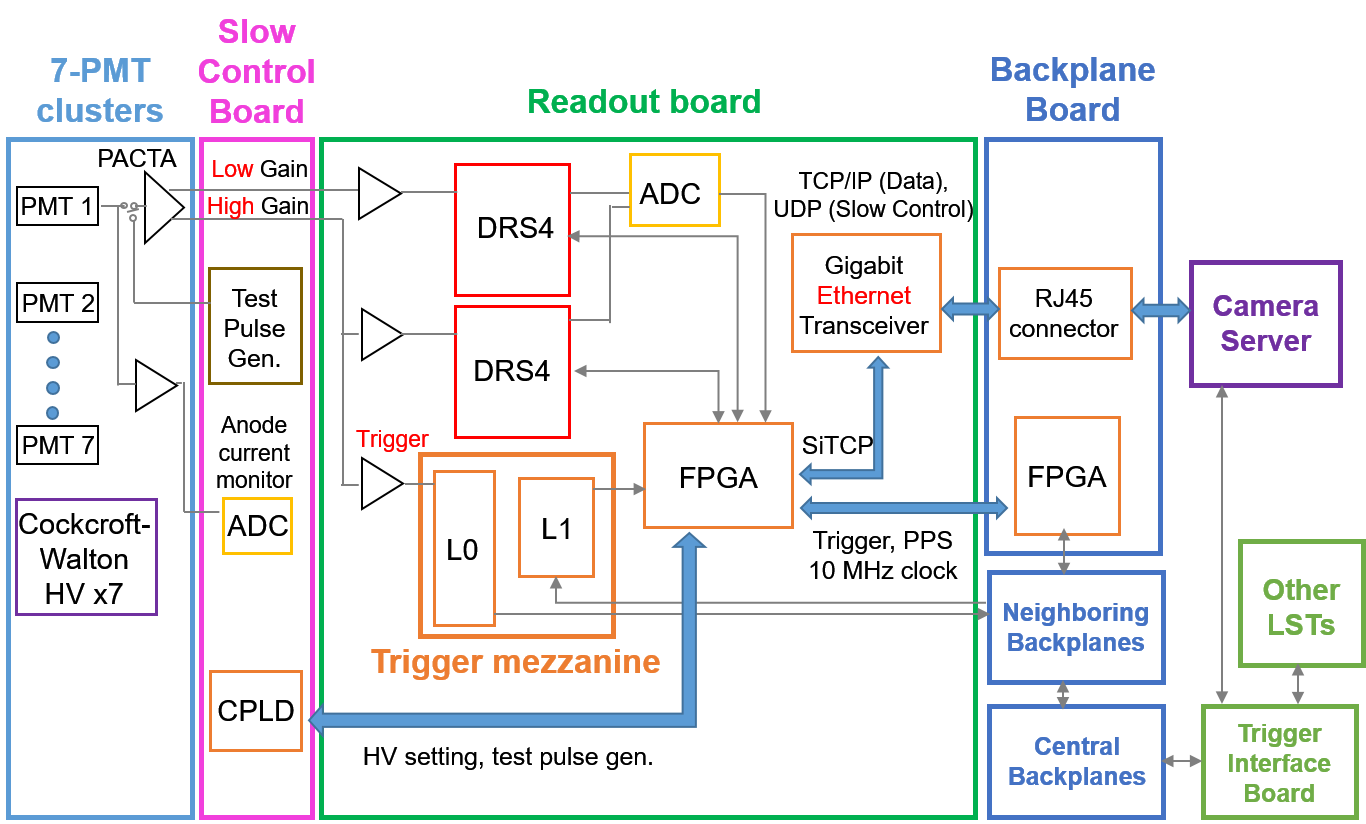}
    \caption{Block diagram of camera components.}
    \label{fig:BlockDiagram}
\end{figure}

Figure \ref{fig:BlockDiagram} shows a block diagram of the readout system. For detecting Cherenkov lights, each module has seven PMTs\cite{Toyama2013ICRC, Toyama2015NIMPA, Mirzoyan2016NIMPA, Mirzoyan2017NIMPA} with light guides. A high voltage for each PMT is generated by the Cockcroft-Walton circuit controlled by the slow control board\cite{Hadasch2015ICRC}. The anode current signal from each PMT is converted to voltage by the current-mode pre-amplifier designed for the CTA\cite{SanuyJInst2012_PACTA}, and connected to the readout board\cite{Kubo2013ICRC,Masuda2015ICRC} based on DRS4 chips.
Those signals are split into three lines with different main amplifiers on this readout board: a high-gain channel, a low-gain channel, and a trigger channel. The high-gain channel is more important to detect dim air showers and achieve lower energy threshold of the LST. In addition to the high-gain channel, it can achieve a wide dynamic range between 0.1 and more than 1000 photoelectrons (p.e.) by using the low-gain channel.

The amplified signals need to be sampled at $\sim$1~GHz to suppress the contamination from the O(ns) night-sky background (NSB) to shower images. A modern flash analog-to-digital converter (FADC) can sample waveforms continuously at GHz rates. However, a FADC is expensive in general and has a high power consumption with large heat dissipation.
Therefore, we adopted DRS4 chips instead of FADCs to achieve the GHz waveform sampling with low power consumption. The DRS4 chips constantly sample the waveform of signals on the low-gain and high-gain lines at 1~GHz frequency. The sampled signal is digitized by an external low-speed analog-to-digital converter (ADC) after the readout board receives the trigger signal.

In parallel, two levels of event triggering are performed with the LST camera. Trigger channel signals are transferred to the analog trigger mezzanine\cite{Gascon2016JInst, Tejedor2013ITNS_L0L1} and combined in each module (level 0). The combined signals are sent to neighboring modules through backplane boards. The signals from three modules are combined to generate a digital trigger pulse with a dedicated threshold for each module (level 1). When the level 1 camera trigger is generated on the trigger mezzanine board, this digital trigger signal is propagated to the central module through backplane boards and sent to the Trigger Interface Board (TIB)\cite{Moya2017ICRC_TIB} connected to the central one. 
The TIB in each telescope has a common timing reference and share the trigger signal among multiple telescopes to generate a stereo trigger in a given small coincidence time window. 
The TIB waits for the $\sim$4~$\mathrm{\mu s}$, which are needed to generate the stereo trigger, then sends the trigger back to the central module. 
This trigger is propagated to all of the modules from the central one. The synchronicity of trigger reception in all modules is asserted by adding individual, calibrated delays to the central trigger signal on each backplane board.

When a field-programmable gate array (FPGA) on the readout board receives the trigger signal, it stops the DRS4 sampling and reads out the stored charges. The analog output from DRS4 is digitized by an external 33~MHz ADC chip with a 12-bit resolution (0.24~mV/ADC). The digitized data blocks are sent to the FPGA and transferred to a data acquisition (DAQ) server through four 10 Gb ethernet links for each telescope by using the SiTCP\cite{Uchida2008_SiTCP} protocol.
The DAQ server receives the data from all of the modules and also other event information such as the trigger and the absolute time from other subsystems, and builds complete single telescope events\cite{Hoffmann2017JPhCS_EVB}. Central CTA DAQ software\cite{Lyard2017ICRC_DAQ} gathers the complete events from the telescopes, then compresses and stores them in ZFITS format\cite{Pence2012_ZFITS}. Those raw data can be analyzed by low-level data processing algorithms for the CTA, {\it ctapipe}\cite{ctapipe}, and a dedicated analysis chain for the LST, {\it lstchain}\cite{lstchain}.

The main calibration tool of the LST-1 is the calibration laser box located at the center of the mirror structure\cite{PalatielloICRC_CaliBox}. The laser generates UV pulses with a pulse width of 400~ps at full width at half maximum, and the laser intensity can be controlled by using neutral density filters. The calibration laser is used for a dedicated calibration run before the standard observations and for taking interleaved calibration events during the observations at a low rate.

\subsection{DRS4}
\label{sec:drs}
The Domino Ring Sampler ASICs have been developed at the Paul Scherrer Institute (PSI), Switzerland, for the MEG experiment\cite{Ritt2010NIMPA}. The latest version, DRS4, is also used in the MAGIC experiment\cite{Aleksi2016_MAGIC_hard}. The DRS4 chip has nine differential input channels with 950~MHz bandwidth. Each channel has 1024 storage capacitors whose write switches are operated via a chain of inverters, called a domino wave circuit. The sampling speed can be changed from 700~MHz up to 5~GHz according to the reference clock from the FPGA. This reference clock is based on an external 10~MHz clock signal propagated from the TIB through the backplane boards. By the common clock, all the DRS4 chips of the LST camera are completely synchronized. 
It is possible to cascade up to eight DRS4 channels for a deep sampling depth. Cascading is realized in the following way; there is an 8-bit register in the chip, each bit of which corresponds to one DRS4 channel. A channel is activated only when the corresponding bit of the register is high. The sampling depth can be summed by connecting the same signal line to multiple channels of DRS4, then activating them one after another. For the LST, four DRS4 channels are cascaded, which means the signal of each input is sampled in 4096 capacitors and the sampling depth is $\sim4~\mathrm{\mu s}$ in the case of 1 GHz sampling. Eight DRS4 chips are mounted on the readout board, and each chip is connected to two PMT channels.
We use a region of interest (ROI) mode to read the stored signals from a part of the capacitors. We set a size of an ROI to 40 capacitors (40~ns) to reduce the dead time.

\section{Calibration method of DRS4 output data}\label{sec:calibration}
The low-level calibration for some intrinsic features of DRS4 is required to correctly estimate the charge value and arrival time of each PMT signal from raw data. Hereafter, we assume that the pulse height of a single photoelectron signal at a high gain channel is about 25~ADC counts (6~mV).

\subsection{Time Lapse Correction and Baseline Subtraction for Individual Capacitors}
\label{subsec:bl_correction}
\begin{figure}
  \begin{minipage}[b]{0.5\linewidth}
    \centering
    \includegraphics[clip, height=6.0cm]{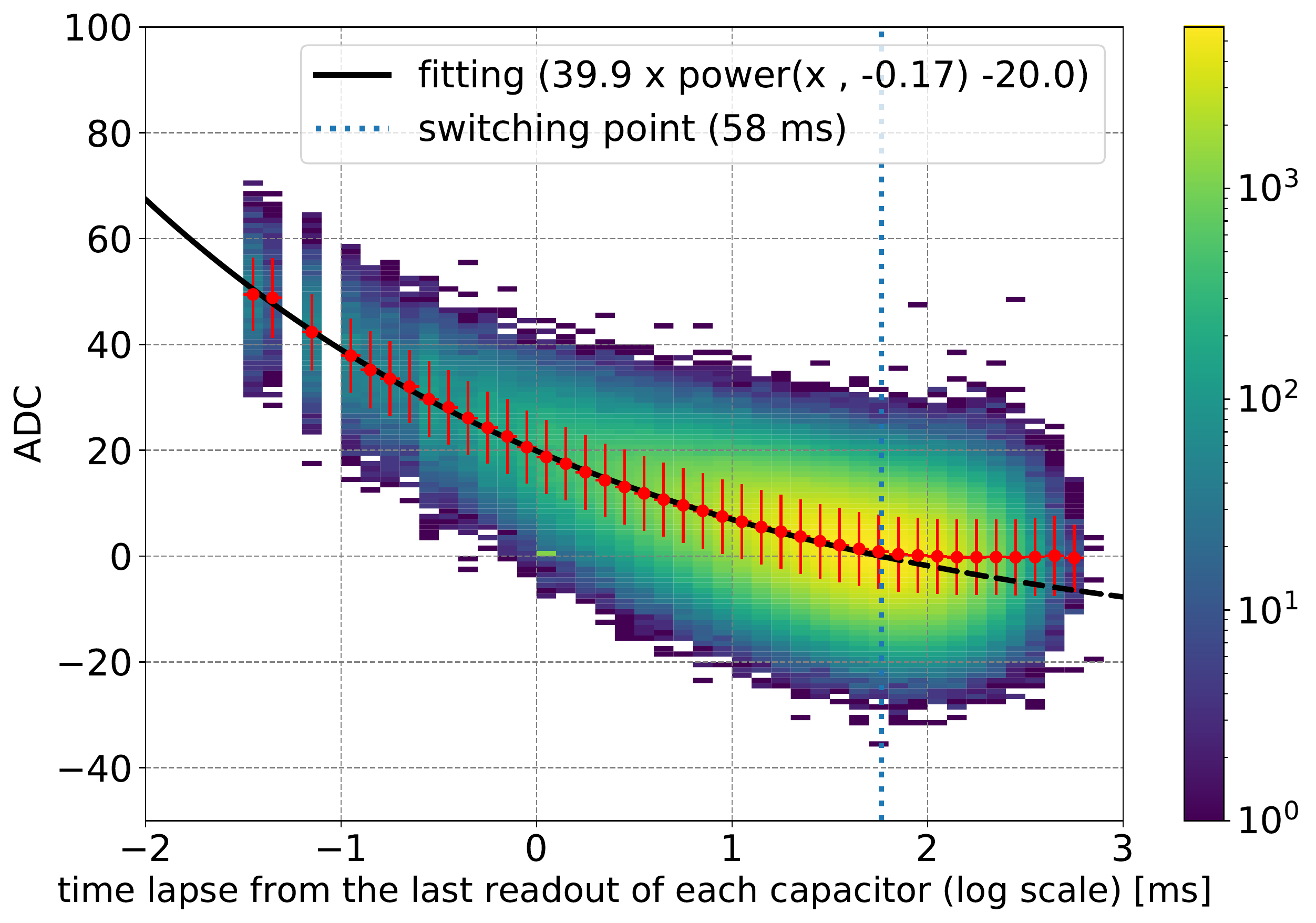}
  \end{minipage}
  \begin{minipage}[b]{0.5\linewidth}
    \centering
    \includegraphics[clip, height=6.0cm]{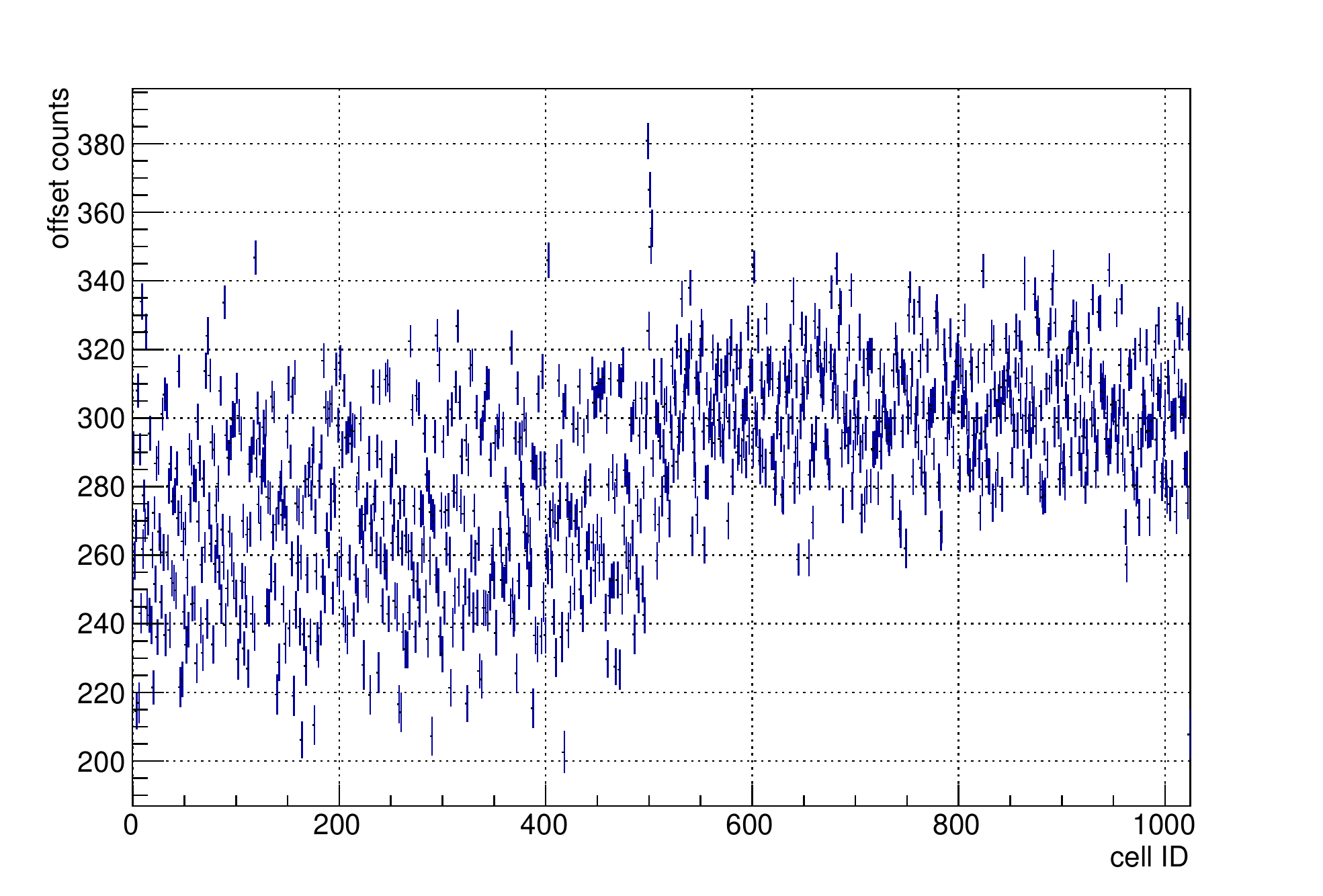}
  \end{minipage}
  \caption{DRS4 pedestal structure for a single DRS4 channel. Left: Pedestal values with different $\Delta T$. Red circles show mean values in each bin with standard deviations. This relation can be described by a power-law function (black solid line) with a smaller $\Delta T$ than 58~ms (blue dotted line). Right: Pedestal value of 1024 capacitors in a single domino chain. Error bars on each point show measurement uncertainties.}
  \label{fig:offset}
\end{figure}

There are two corrections to calibrate the DRS4 pedestal values: a time-lapse correction and a cell-wise baseline correction. The pedestal values are related to the time lapse from the last readout of each capacitor ($\Delta T$). Figure \ref{fig:offset} (left) shows a relation between $\Delta T$ and pedestal values. It is known that this relation can be well fitted by a power-law function \cite{Sitarek2013NIMPA}. Each DRS4 chip has common fitting parameters to all capacitors, and those parameters mildly depend on individual DRS4 chips. We can compensate for this effect using this fitting function. 
This feature disappears and the time lapse correction is not needed when $\Delta T$ is larger than around 60~ms.
To find the cell-wise pedestal, it is needed to apply this correction with nominal parameters to remove the $\Delta T$ dependence or to use only data with larger $\Delta T$. Figure \ref{fig:offset} (right) shows the pedestal values of each capacitor in a single domino chain (1024 capacitors). It is well known that there is a step structure in the distribution, with the 511th cell as a boundary. The differences of those pedestal values between the capacitors are much larger than the noise in a single capacitor.

\begin{figure}
  \begin{minipage}[b]{0.5\linewidth}
    \centering
    \includegraphics[clip, height=6.0cm]{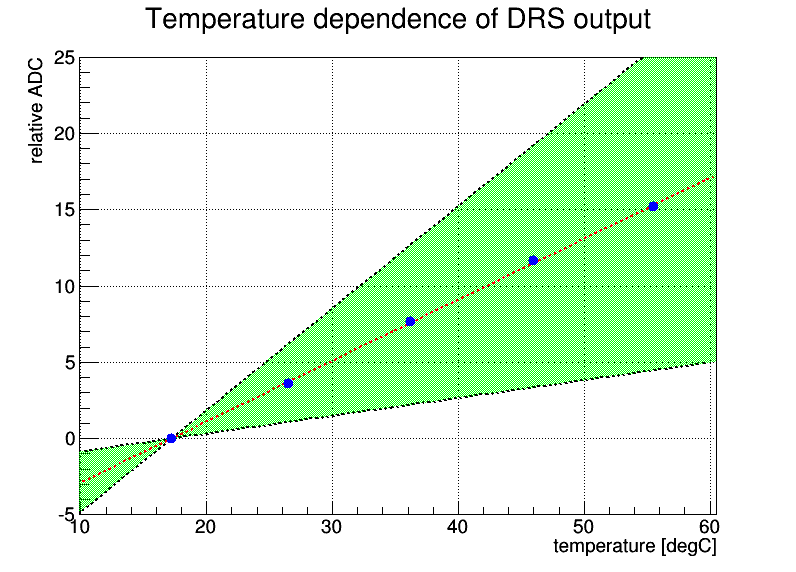}
  \end{minipage}
  \begin{minipage}[b]{0.5\linewidth}
    \centering
    \includegraphics[clip, height=6.0cm]{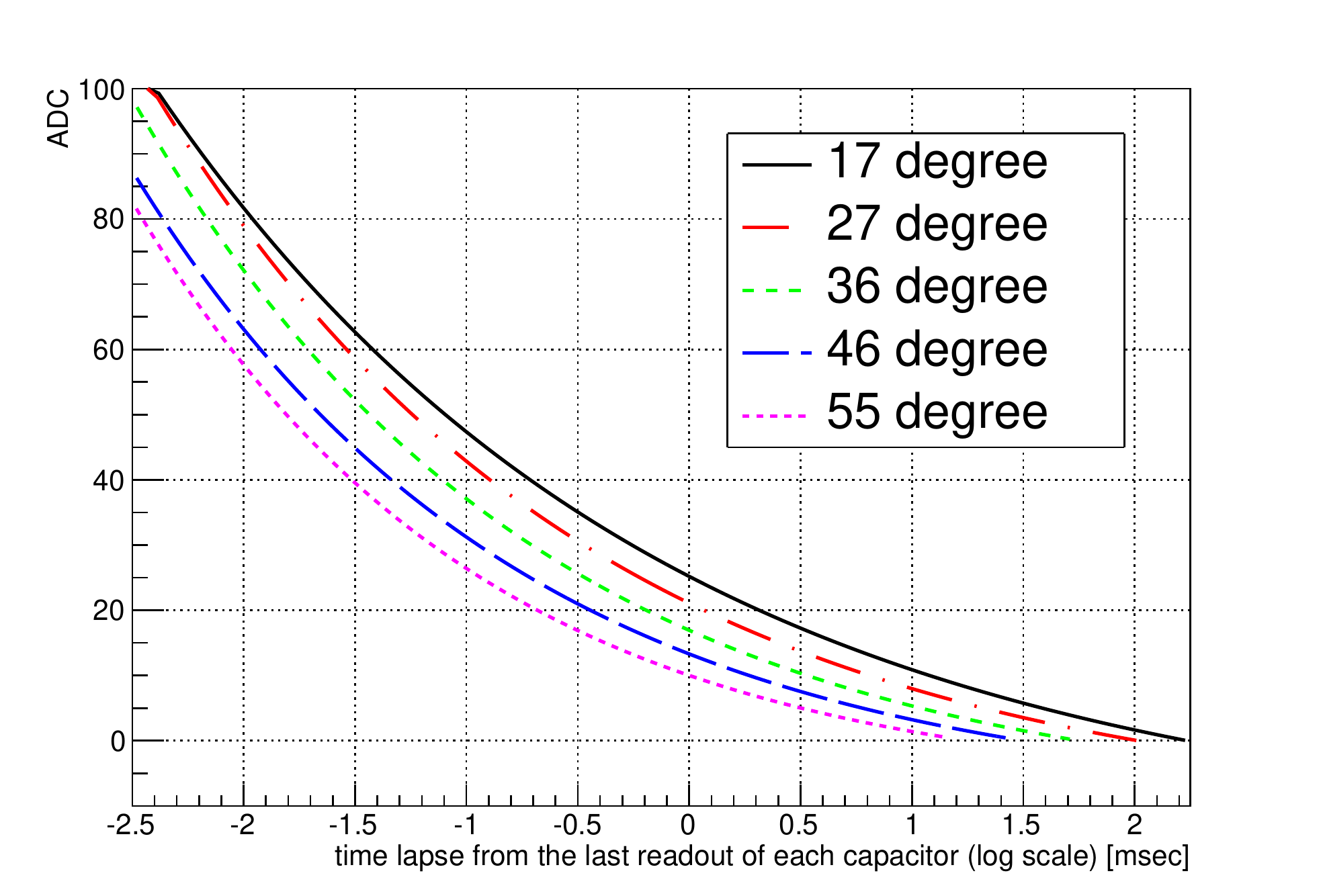}
  \end{minipage}
  \caption{Temperature effect on the parameters of the DRS4 pedestal. Left: Relative pedestal value with respect to the pedestal at 17~$\mathrm{{}^{\circ}C}$ as a function of temperature. This pedestal values are computed after the time lapse correction. Right: DRS4 pedestal as a function of $\Delta T$ at different temperatures.}
  \label{fig:offset_temperature_dep}
\end{figure}

Those parameters depend on the temperature. Figure \ref{fig:offset_temperature_dep} (left) shows a relation between temperature and relative cell-wise pedestal values. The temperature in a chamber is controlled from 10$\mathrm{{}^{\circ}C}$ to 50$\mathrm{{}^{\circ}C}$ at laboratory and measured with a sensor on the readout board. The mean gradient with respect to the temperature is 0.5 $\mathrm{{ADC}/{{}^{\circ}C}}$. The cooling system can maintain the temperature inside the camera within $\pm$5$\mathrm{{}^{\circ}C}$, which allows the
pedestal values to be kept stable at $\pm$0.6~mV, corresponding to $\sim$0.1 p.e. at a high gain channel. Figure \ref{fig:offset_temperature_dep} (right) shows $\Delta T$ curves for different temperatures. Those curves approach zero with larger $\Delta T$ at higher temperatures.

To prepare the parameters of the fitting function for the time lapse correction and cell-wise pedestal, a dedicated pedestal data with a random Poisson trigger is needed. 
If triggers are generated at a constant frequency, it causes structures in the last reading time of individual capacitors due to the folding of the frequency of triggers with the sampling frequency of DRS4 chips (beating effects). On the other hand, random triggers could give more similar conditions to what happens during the standard observations when the shower triggers are generated following Poissonian statistics.
A random trigger can be generated in FPGA on the readout boards based on random numbers generated by a linear feedback shift register. The frequency of the random trigger can be controlled by slow control commands, and 2~kHz is used for the pedestal run to cover various $\Delta T$. This dedicated pedestal run is performed daily after powering up the camera, when temperature is sufficiently stable.


\subsection{Spike Correction}
\label{subsec:spike_correction}
\begin{figure}
  \begin{minipage}[b]{0.5\linewidth}
    \centering
    \includegraphics[clip, height=6.0cm]{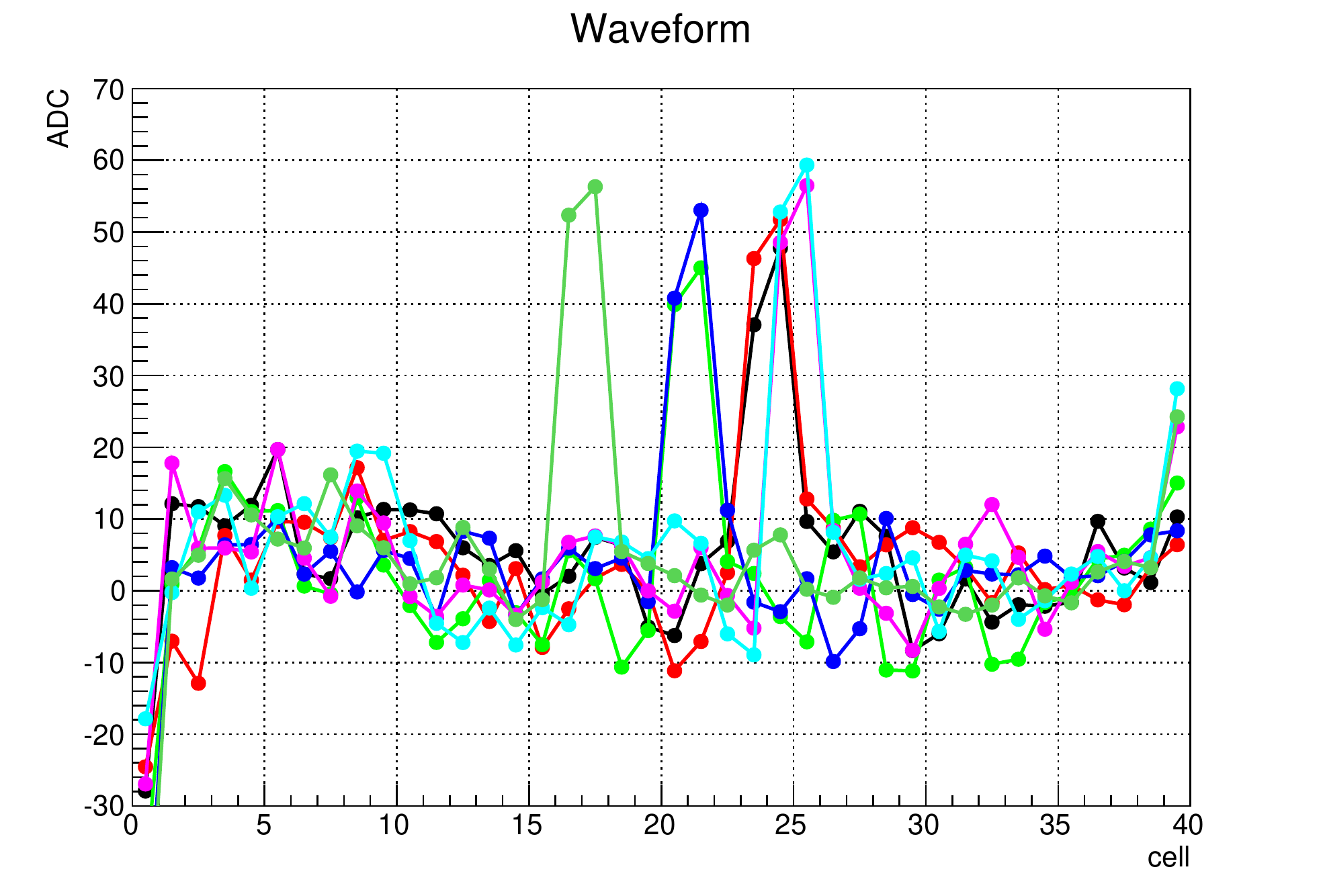}
  \end{minipage}
  \begin{minipage}[b]{0.5\linewidth}
    \centering
    \includegraphics[clip, height=6.0cm]{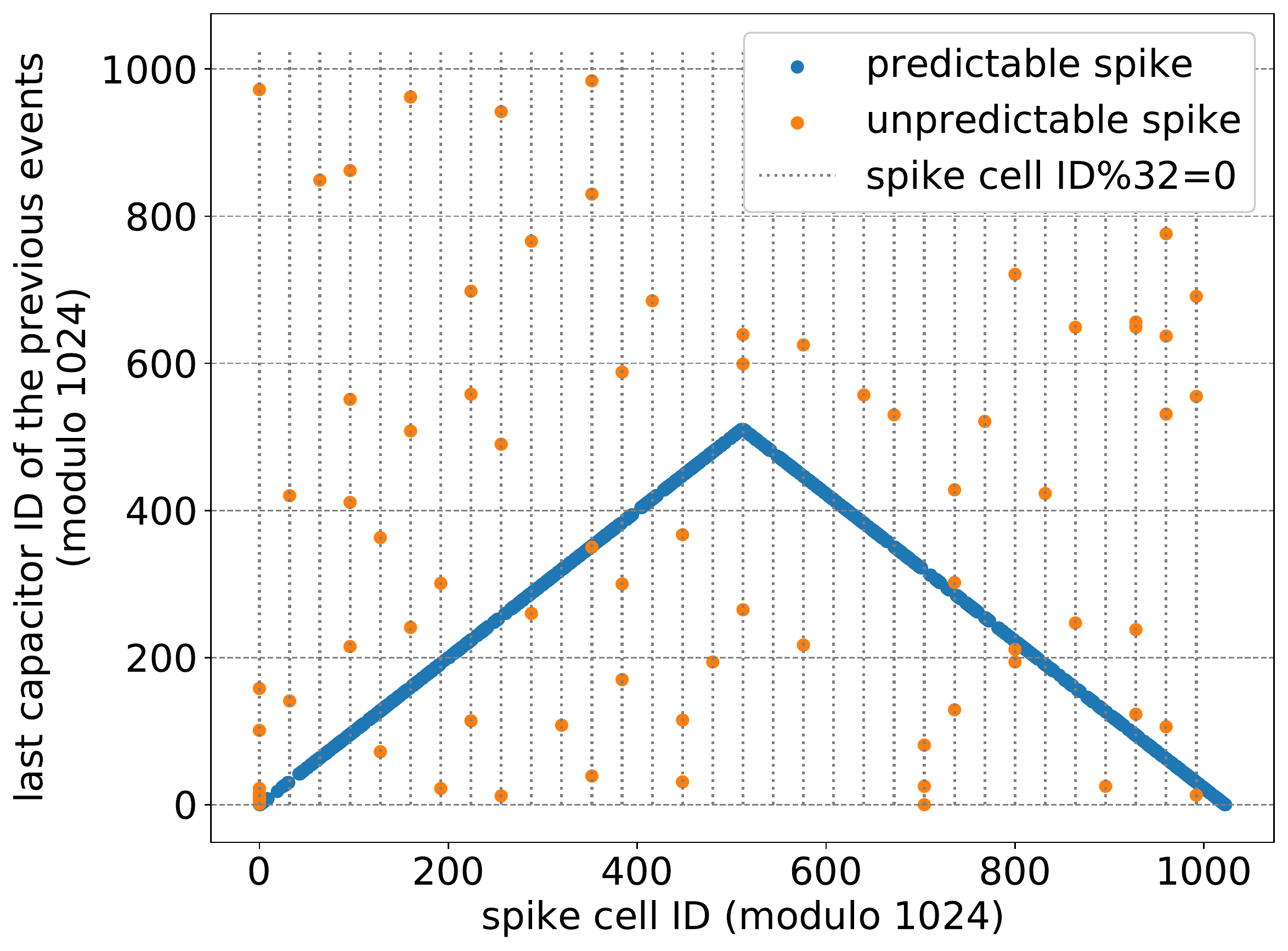}
  \end{minipage}
  \caption{Spike features of the DRS4 chips. Left: Waveform of seven channels including spike cells. Right: Relation between the spike cell ID and the last capacitor ID in the readout windows of the previous events. Those cell IDs are computed by an operation of modulo 1024. There are predictable consecutive spikes (blue) and unpredictable ones (orange) at every 32nd capacitor (gray dotted lines).}
  \label{fig:spike}
\end{figure}

Two consecutive cells in ROI have a spike feature under a certain condition as shown in Figure \ref{fig:spike} (left). The spike height is always 40--50 ADC counts, corresponding to 1--2 p.e.. The position of the spike depends on the last capacitors recording signal in the previous event. Although the reason for this feature is not interpreted perfectly, it is suggested that those spikes could be caused by remaining read bits at the previous readout. Figure \ref{fig:spike} (right) shows the relation between the position of spike cells and the last capacitors recording signal in the previous readout window. The spike cells are defined as ones which have higher ADC counts than 15 ADC in this figure. Since this relation is predictable, and these features appear without exceptions, we can compensate for this feature by interpolating between neighbor cells or subtracting ADC counts corresponding to the pulse height. DRS4 chips on all of the modules are running based on the common clock reference, and domino waves of all DRS4 chips are basically synchronized as described in Section \ref{sec:drs}. Thus, a similar region of capacitors in the domino wave of all pixels is read out for every event. It is possible that most of the pixels have such spike cells in the readout window when the conditions are fulfilled. 

In this Figure \ref{fig:spike} (right), a periodic pattern is also seen at every 32nd capacitor, which was already known in the MAGIC experiment\cite{Sitarek2013NIMPA}. 
This feature seems to be related to the structure of DRS4 chips, which is segmented into 32 parts with 32 capacitors for each channel. 
Those capacitors sometimes have larger ADC counts. Though it is difficult to predict the condition and correct it, we revealed that this effect is shown only when $\Delta T$ is larger than about 100~ms. This effect is therefore negligible at a standard trigger rate (9--10~kHz) for the observation.

\subsection{Pulse Peak Time Correction}
\label{sec:time_correction_peak}
A sampling frequency in a DRS4 chip is not perfectly constant. Each capacitor has a cell-wise fixed sampling interval in addition to a small random jitter. This interval varies from one capacitor to another by about 10\%. 
Basically, a timing calibration of switched capacitor arrays to obtain the individual sampling intervals needs a reference signal such as a sine wave \cite{Shaver2014ITNS_DRS_timecal}. However, there is no such equipment on the readout boards. We have thus developed alternative methods for a pulse peak time correction and an integrated charge correction individually.

\begin{figure}
    \centering
    \includegraphics[height=8cm]{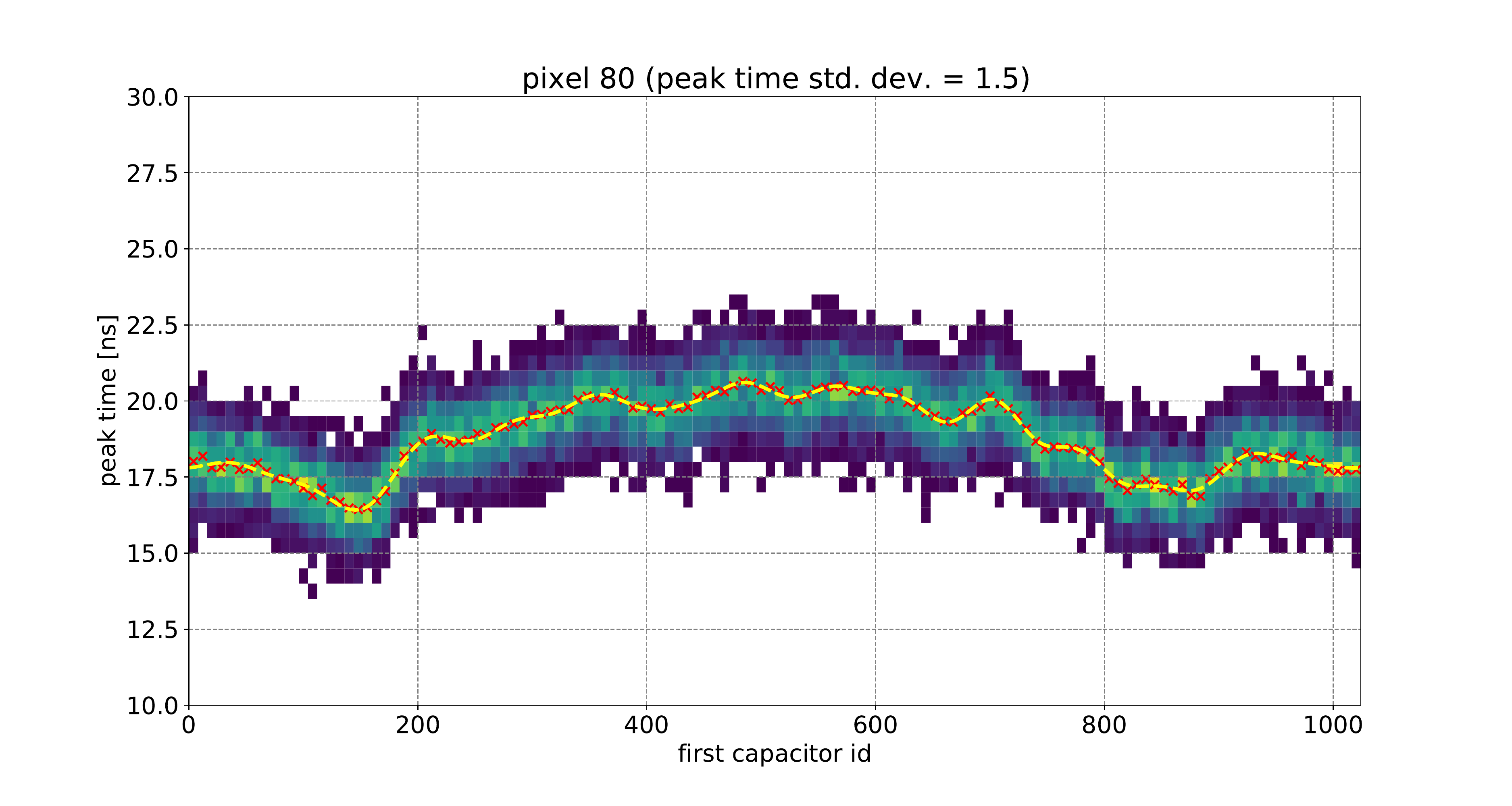}
    \caption{Relation between the pulse peak time and the first capacitor ID in a readout window of a given pixel. Red crosses show the mean peak time in each binning, and those data points are expanded into a Fourier series (yellow line).}
    \label{fig:time_correction}
\end{figure}

If we assume constant sampling frequency contrary to the fact, a pulse peak time depends on a readout position in the domino wave. Figure \ref{fig:time_correction} shows a relation between the first capacitor id of the readout window and the pulse peak time with the assumption of constant sampling intervals.
Since this relation is common for each channel on the same DRS4 chip, the first capacitor id is folded into modulo 1024.
The dispersion of the pulse peak time depending on the first capacitor corresponds to integrated sampling intervals at each part of the domino ring between the first capacitor in the readout window and pulse peak position.
In the case of calibration-laser events used for this method, the laser signal is located at the center of the readout window.
Thus, we can correct these sampling intervals of $\sim$20 capacitors by using a Fourier series expanded from this relation\cite{Sitarek2013NIMPA}.
This relation depends on the pulse position in the readout window because a number of capacitors in front of the pulse peak is different. Therefore, we are currently developing a method to correct the pulse peak time independent of the pulse position in the window.


\subsection{Integrated Charge Correction}
\label{sec:time_correction_charge}
\begin{figure}
    \centering
    \includegraphics[height=7cm]{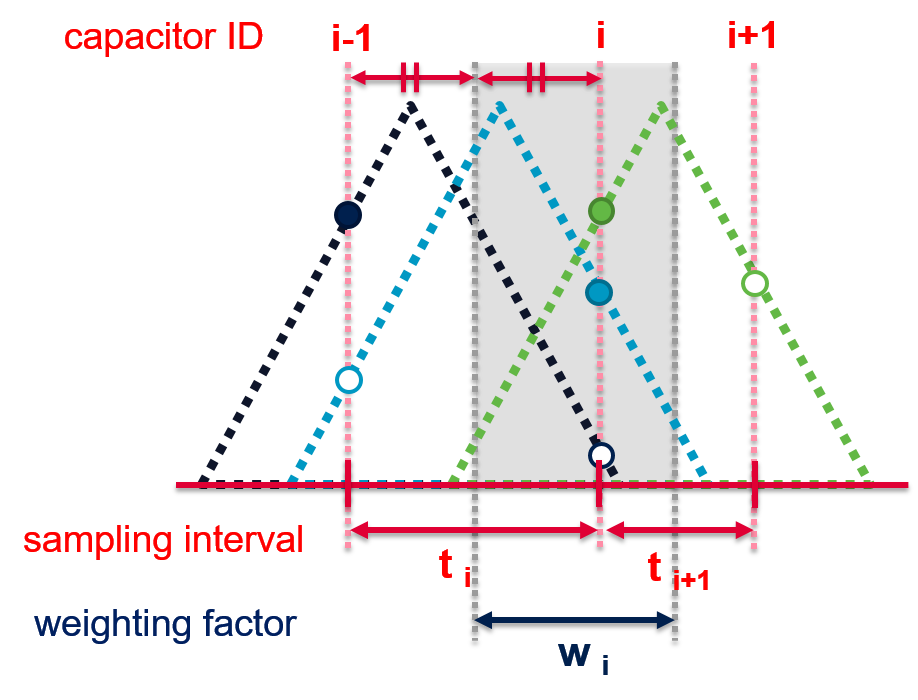}
    \caption{Concept of the method to find the weighting value with test pulse. Three test pulses are shown as dotted lines (black, blue, green), and sampled data points at each capacitor are shown as circles. Peak points of each waveform are indicated by filled circles, and others are indicated by open circles. In the ideal case, the i-th capacitor have peak points when test pulse peaks are located between a middle of sampling interval of the i-th capacitor and one of the neighboring capacitor (gray shadow region). In this figure, such cases correspond to blue and green pulses. The probability that the i-th capacitor have peak points corresponds to a ratio of gray shadow region to sum of sampling intervals of all capacitors.}
    \label{fig:time_cal_pulse_method}
\end{figure}

The sampling inhomogeneity described in the previous section also affects the integrated charge, which is the product of ADC counts and time. 
Instead of using the individual sampling intervals, We use a method using test pulses to find the weighting factor for the charge integration instead of the sampling intervals. Figure \ref{fig:time_cal_pulse_method} shows a concept of the method to obtain the weighting factor. Firstly, we need test pulse data with large statistics (one million of events allows the accuracy of $\sim$3\%). Those test pulses must be driven by clocks which are different from the one used for DRS4, in order to avoid synchronisation between test pulses and domino waves, sweeping uniformly the relative time between both. With those test pulse events, we count the number of events for each capacitor that has a peak. In the ideal case, the i-th capacitor has a peak when the actual pulse peak is located between a middle of sampling interval of the i-th capacitor and one of the neighboring capacitor. We will use the ratio of this number to the total number of events as a weighting factor. When we define this weighting factor as $w_i$ and sampling interval as $t_i$ of the i-th capacitor, those relations are explained by the Equation (\ref{eq:weight}).
\begin{eqnarray}
w_{i} &=& \cfrac{t_i + t_{i+1}}{2} \nonumber\\
&=& \cfrac{\text{a number of events with a peak at the i-th capacitor}}{\text{a total number of events}} \times 1024~\mathrm{ns}
\label{eq:weight}
\end{eqnarray}
To compensate for the uneven sampling time intervals, we have only to integrate the product of ADC counts and the weighting factor for each capacitor. This procedure corresponds to a trapezoidal integration of waveforms as shown in the Equation (\ref{eq:charge_integ}). 
\begin{eqnarray}
Charge = \sum_{i=n}^{n+m} \cfrac{ADC_i + ADC_{i+1}}{2} \times t_{i+1} \sim \sum_{i=n}^{n+m} \cfrac{t_i + t_{i+1}}{2} \times ADC_{i+1} = \sum_{i=n}^{n+m} w_i \times ADC_i
\label{eq:charge_integ}
\end{eqnarray}
This inhomogeneity does not have a large effect for small numbers of photoelectrons because the statistical uncertainty is much larger than one from this effect. On the other hand, this correction is important to evaluate the charge of a large signal, especially above a thousand photoelectrons. Since it strongly affects the spread of the charge distribution of such signals, it can bias the excess noise factor method\cite{Gaug2005ICRC_ENF}.


\section{Readout performance}
\label{sec:result}

\subsection{Noise Level}
\label{sec:noise}
\begin{figure}
  \begin{minipage}[b]{0.5\linewidth}
    \centering
    \includegraphics[clip,height=7.0cm]{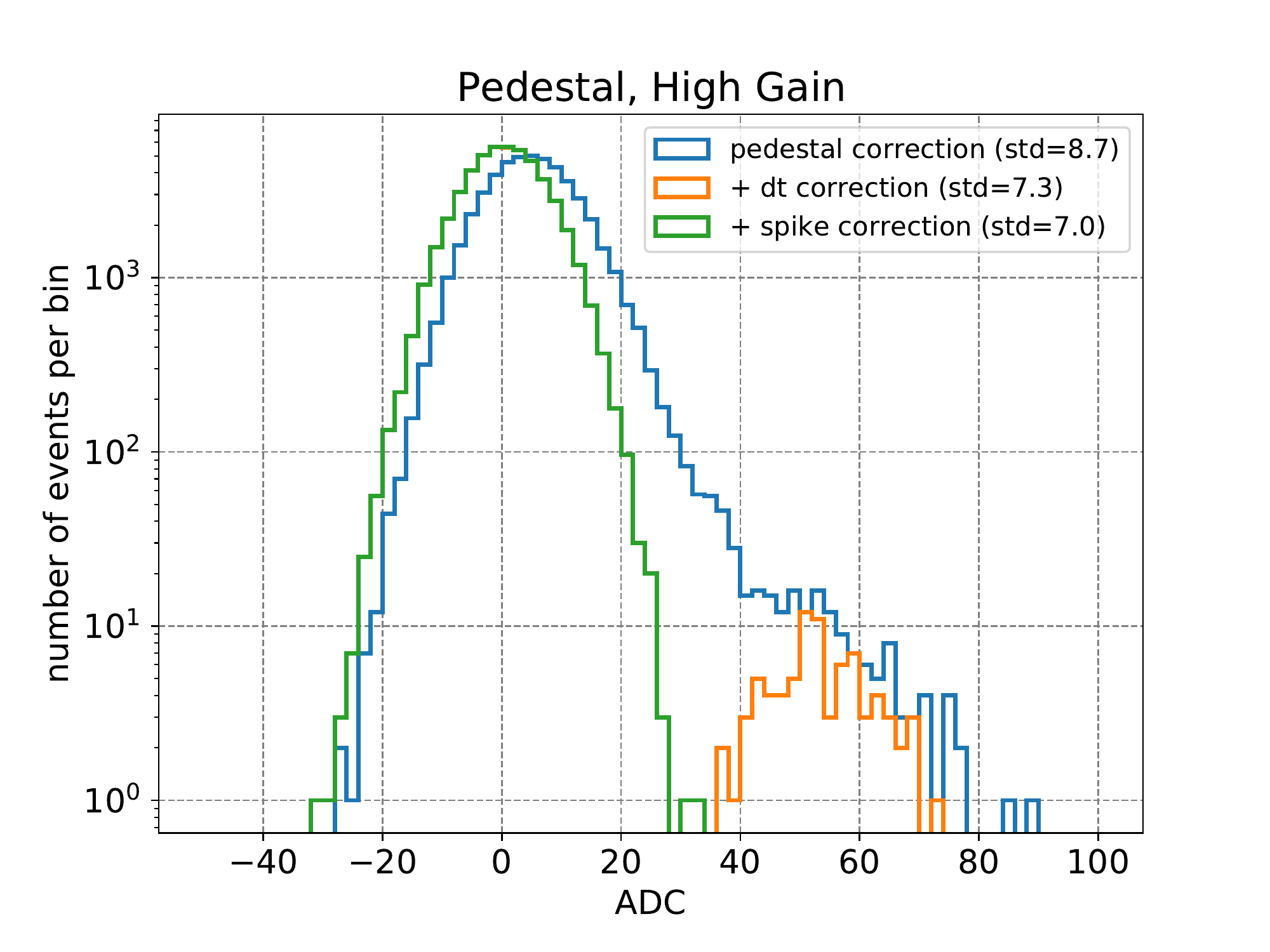}
  \end{minipage}
  \begin{minipage}[b]{0.5\linewidth}
    \centering
    \includegraphics[clip,height=7.0cm]{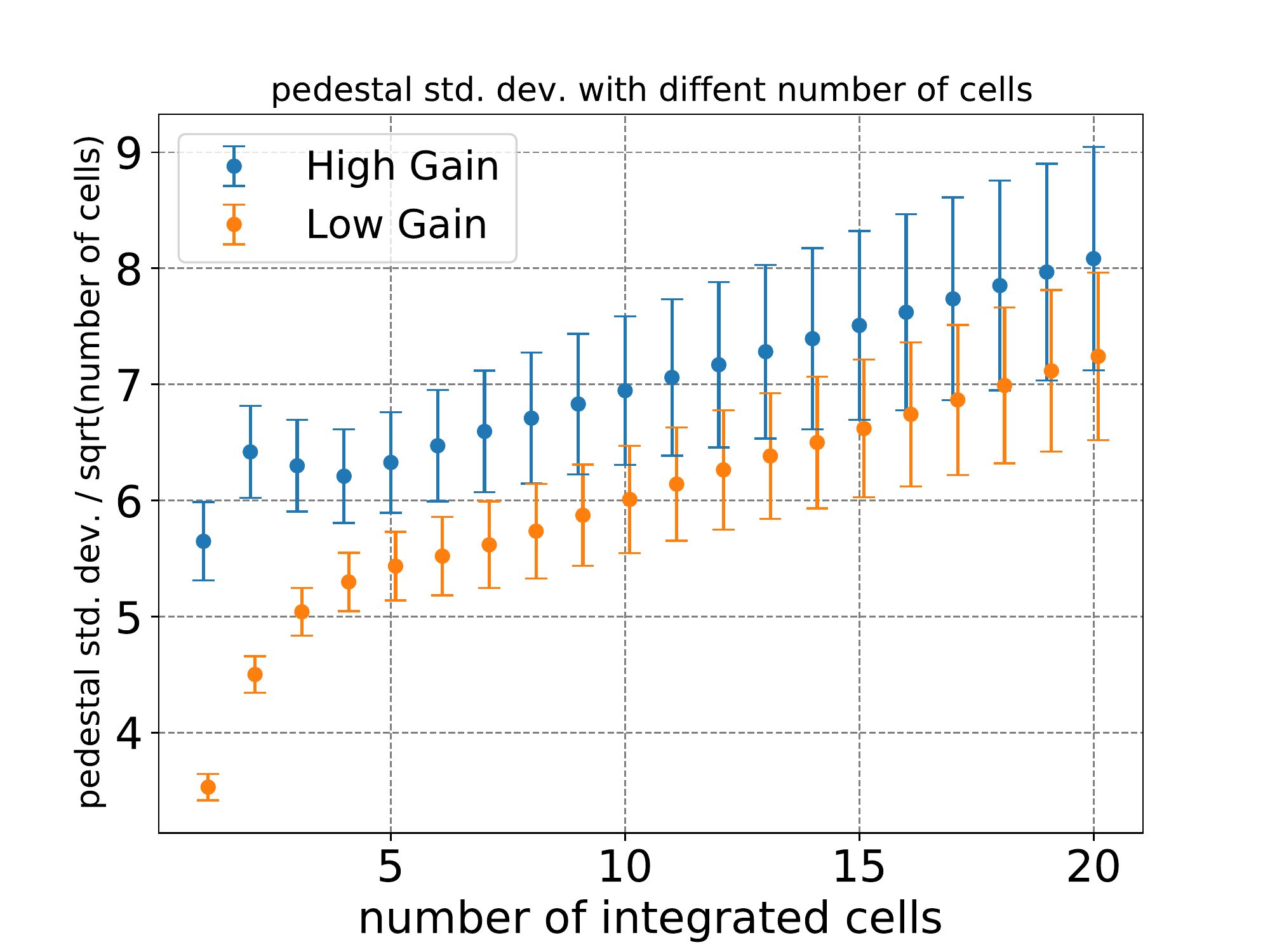}
  \end{minipage}
  \caption{Left: A pedestal distribution with a unit of ADC counts at each cell after each step of the corrections: pedestal subtraction (blue), $\Delta T$ correction (orange), and spike correction (green). Right: Standard deviations of the pedestal by integrating different numbers of cells for the high gain (blue) and the low gain (orange) channels. Error bars on each point show the standard deviation of this parameter among all 1855 pixels in the LST-1 camera.}
  \label{fig:ped_hvoff}
\end{figure}

We evaluated a standard deviation of pedestal value as an index of a noise level after the dedicated corrections described in Section \ref{sec:calibration}. Figure \ref{fig:ped_hvoff} (left) shows the distribution of pedestal ADC at each cell after each correction step. Each step of the correction works well to compensate each component, and the standard deviation of the pedestal at the high gain channel is 7.0~ADC counts (0.28 p.e.) after all corrections. Since the charge value is derived from the integration of ADC values among multiple cells, we investigated the effect of a number of integrated cells on the standard deviation of the pedestal. Figure \ref{fig:ped_hvoff} (right) shows the standard deviation of the pedestal integrated with different numbers of cells. Here, we evaluated the standard deviation of the pedestal as one divided by a square root of the number of integrated cells assuming that the noise is not correlated. If noise components are random, this parameter should be the same with a different number of integrated cells. However, this parameter is higher with more integrated cells. A possible reason is that some noise components have a lower frequency than the charge sampling rate. The eight-cell integration is used to derive the charge of Cherenkov signal for the standard analysis. Hereafter, the eight-cell integration is used to evaluate a noise level.

\begin{figure}
  \begin{minipage}[b]{0.5\linewidth}
    \centering
    \includegraphics[clip,height=7.0cm]{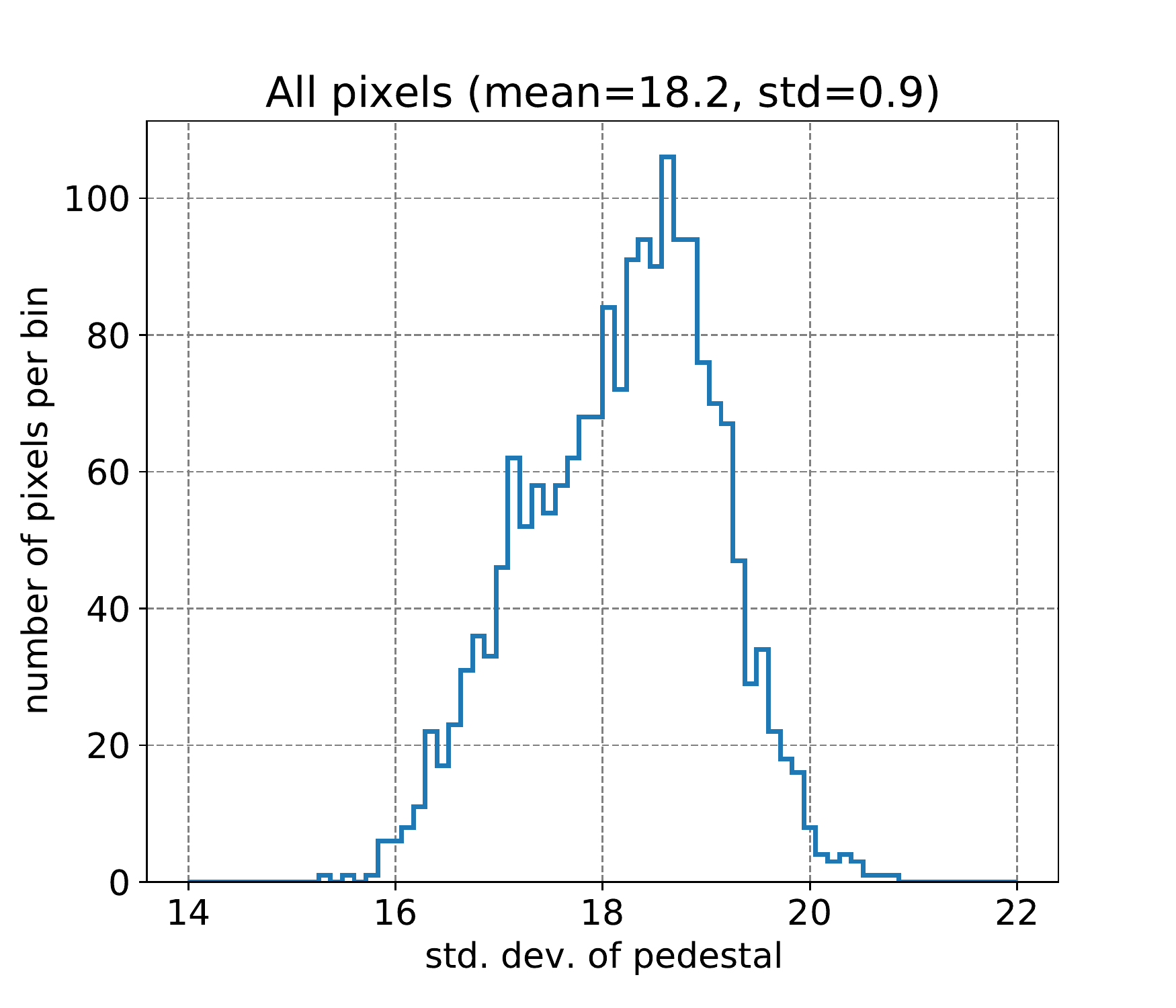}
  \end{minipage}
  \begin{minipage}[b]{0.5\linewidth}
    \centering
    \includegraphics[clip,height=7.0cm]{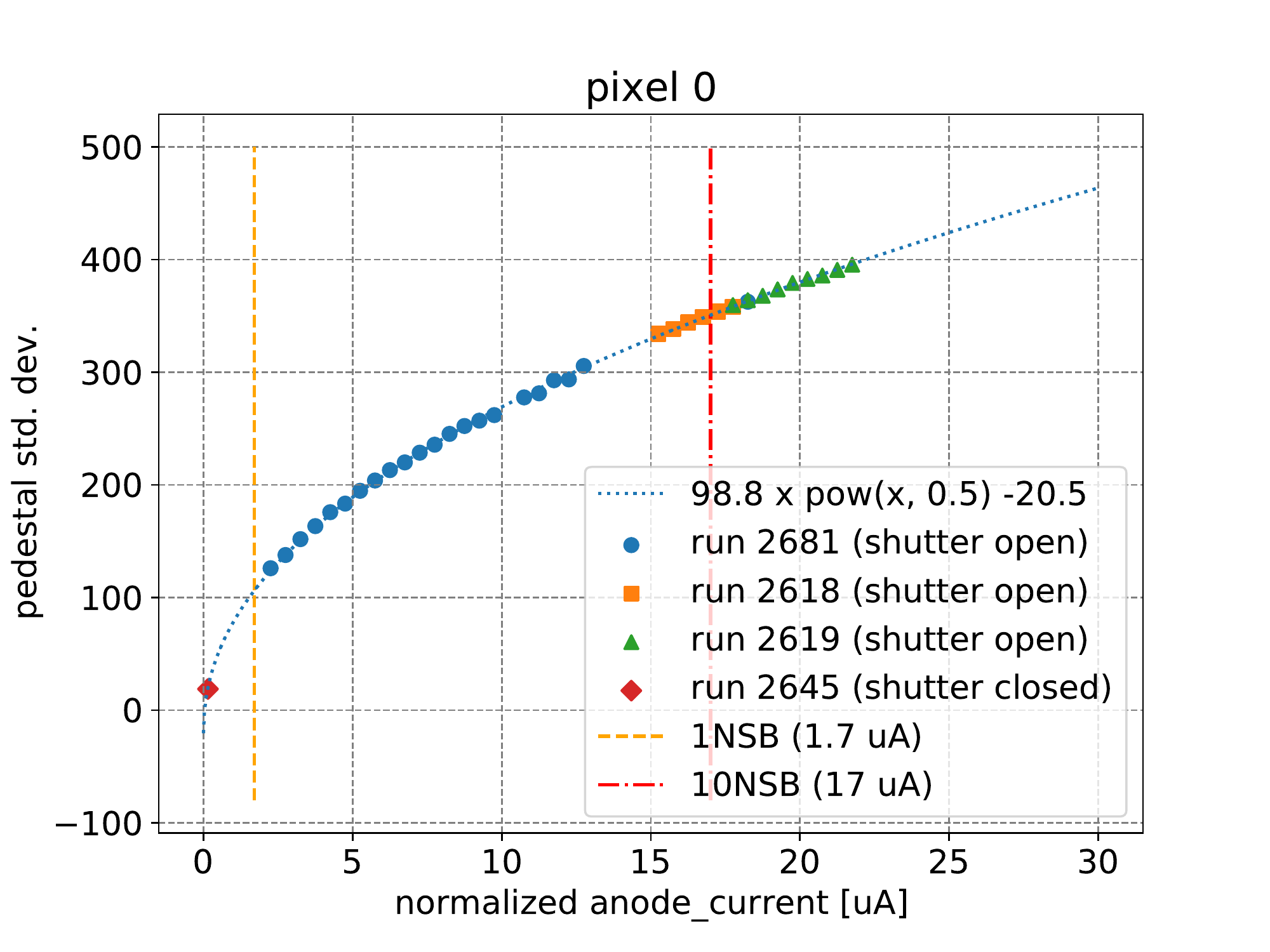}
  \end{minipage}
  \caption{Left: Standard deviation of the pedestal with the eight-cell integration among all camera pixels without a high voltage applied to PMTs. Right: Standard deviation of the pedestal with the eight-cell integration under different NSB levels as a function of anode current divided by attenuation factors, corresponding to the NSB level. An orange dashed line shows 1~NSB level (1.7~$\mathrm{\mu A}$), and a red dashed-dotted line shows 10~NSB level(17~$\mathrm{\mu A}$). Data taken with the shutter open and a high voltage applied to PMTs is shown with blue circle, orange square, and green triangle markers. A red diamond shows data taken with the shutter closed and a high voltage applied to PMTs. A blue dotted line shows a fitting line with a power-law function with an offset.}
  \label{fig:ped_hv}
\end{figure}

Figure \ref{fig:ped_hv} (left) shows a distribution of pedestal value with the eight-cells integration among all pixels in the camera without a high voltage applied to PMTs. The mean value is 18.2~ADC counts, which corresponds to 0.21~p.e., assuming that a charge value of a single photoelectron signal is 80--90~ADC counts. This noise component is contributed only by electronics. On the other hand, background light are the dominating source of noise during standard observations. 
Figure \ref{fig:ped_hv} (right) shows the standard deviation of the pedestal with a high voltage applied to PMTs under the different NSB levels. Due to three types of attenuation factors, which are different pixel by pixel, we used the normalized anode current divided by the attenuation factor of each pixel to evaluate the relationship with the same parameters for all pixels. A fitting function indicates that this relation is well described by a square-root function with an offset. The fitting parameters are almost the same among all pixels. This relation holds even at 10 times higher than the dark NSB level.

\subsection{Charge Uniformity}
\begin{figure}
    \centering
    \includegraphics[height=8cm]{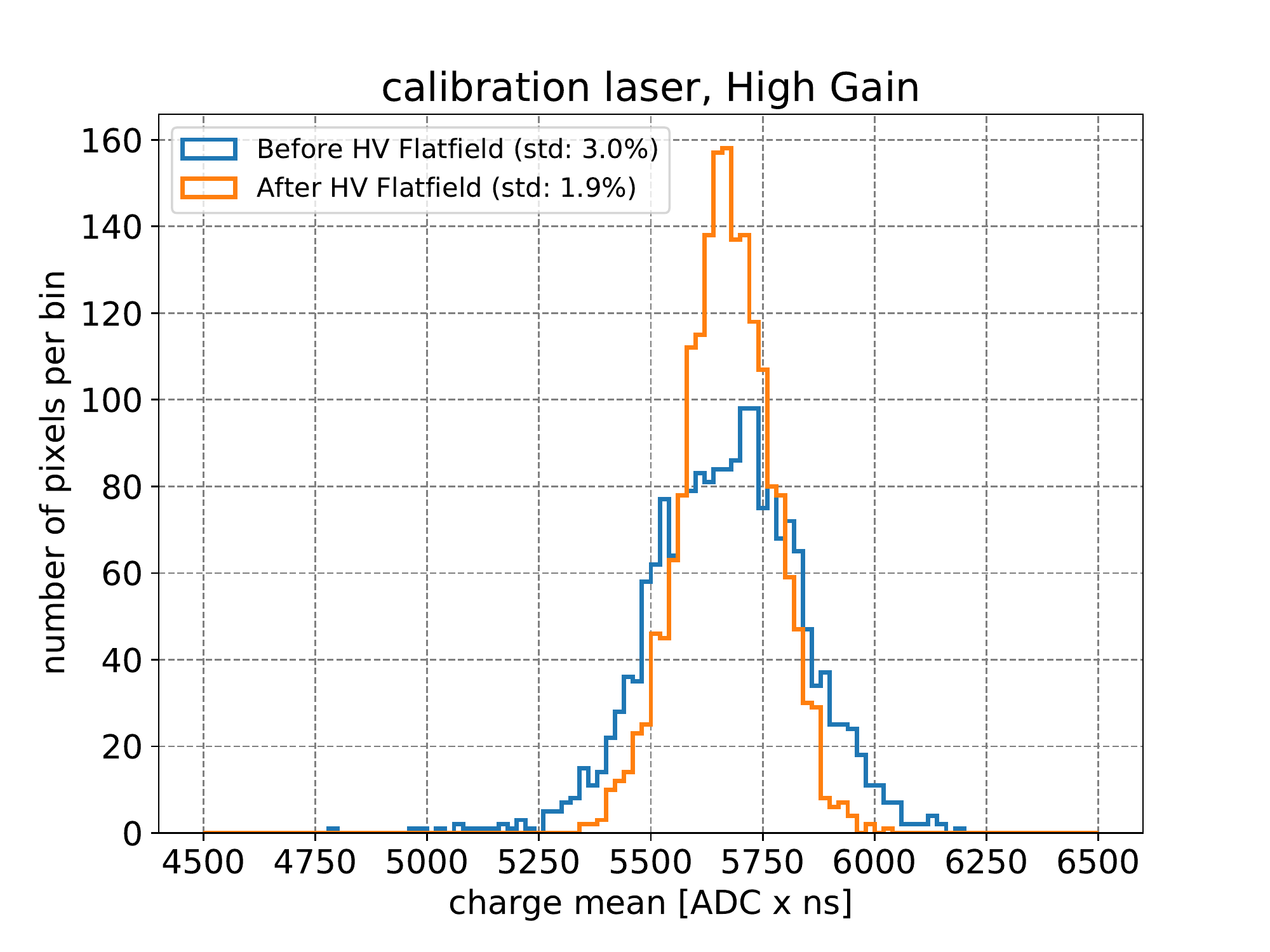}
    \caption{Mean charge value with calibration laser data before the high-voltage flat-fielding (blue) and after the high-voltage flat-fielding (orange).}
    \label{fig:charge_dist}
\end{figure}

For a homogeneous response of the camera, it is needed to adjust high voltages for all pixels to get the same amount of output charges from PMTs against the calibration laser. This procedure is called high-voltage flat-fielding. 
Figure \ref{fig:charge_dist} shows the charge distribution at a high-gain channel before and after the high-voltage flat-fielding in the same night in August 2020. The standard deviation of this distribution with the calibration laser was improved by the high-voltage flat-fielding from 3.0\% to 1.9\%. The previous high-voltage flat-fielding was performed in January 2020 and the PMT gain was varied after that time.

\subsection{Charge Resolution}
\begin{figure}
  \begin{minipage}[b]{0.5\linewidth}
    \centering
    \includegraphics[clip, height=7.0cm]{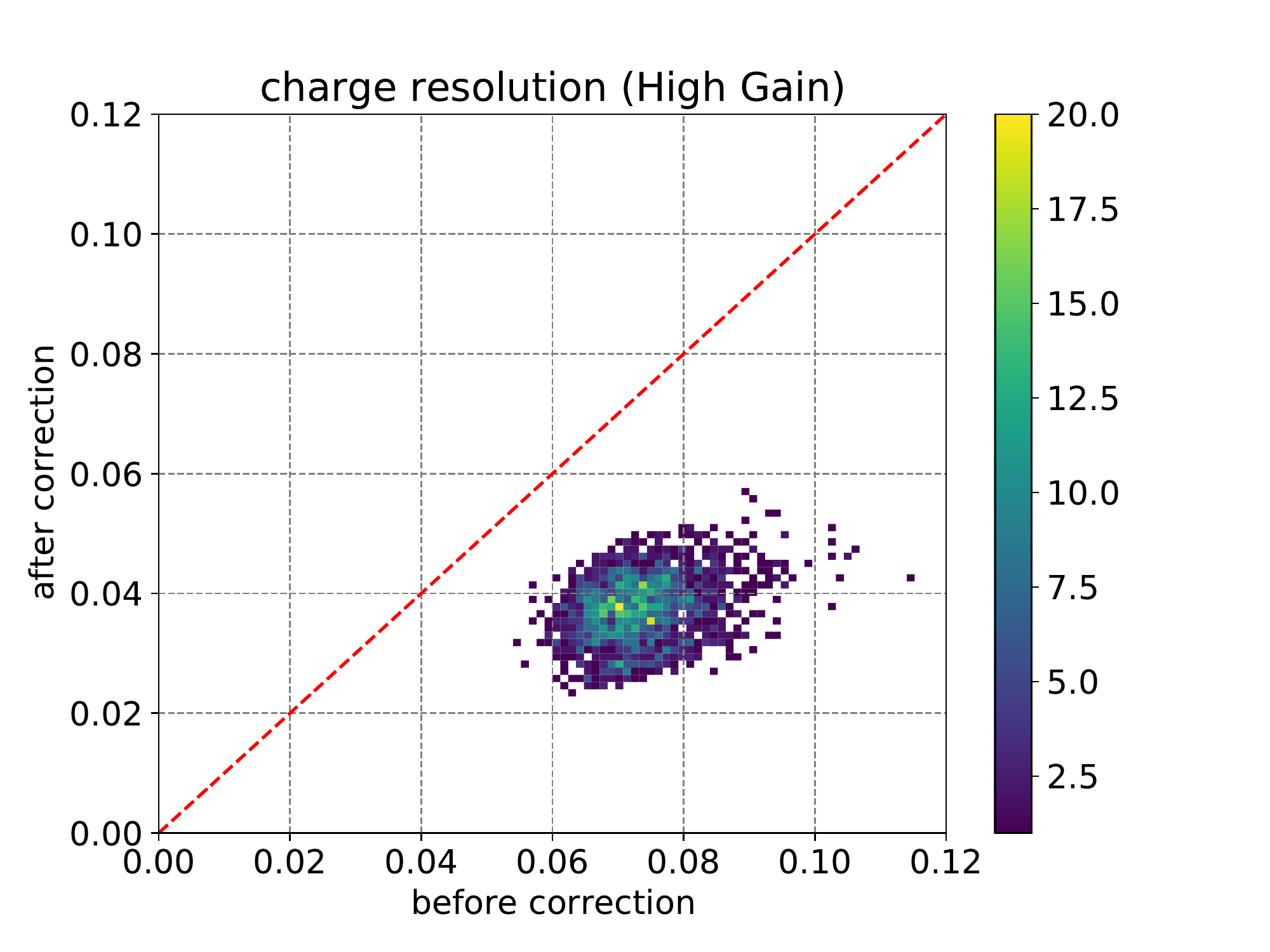}
  \end{minipage}
  \begin{minipage}[b]{0.5\linewidth}
    \centering
    \includegraphics[clip, height=7.0cm]{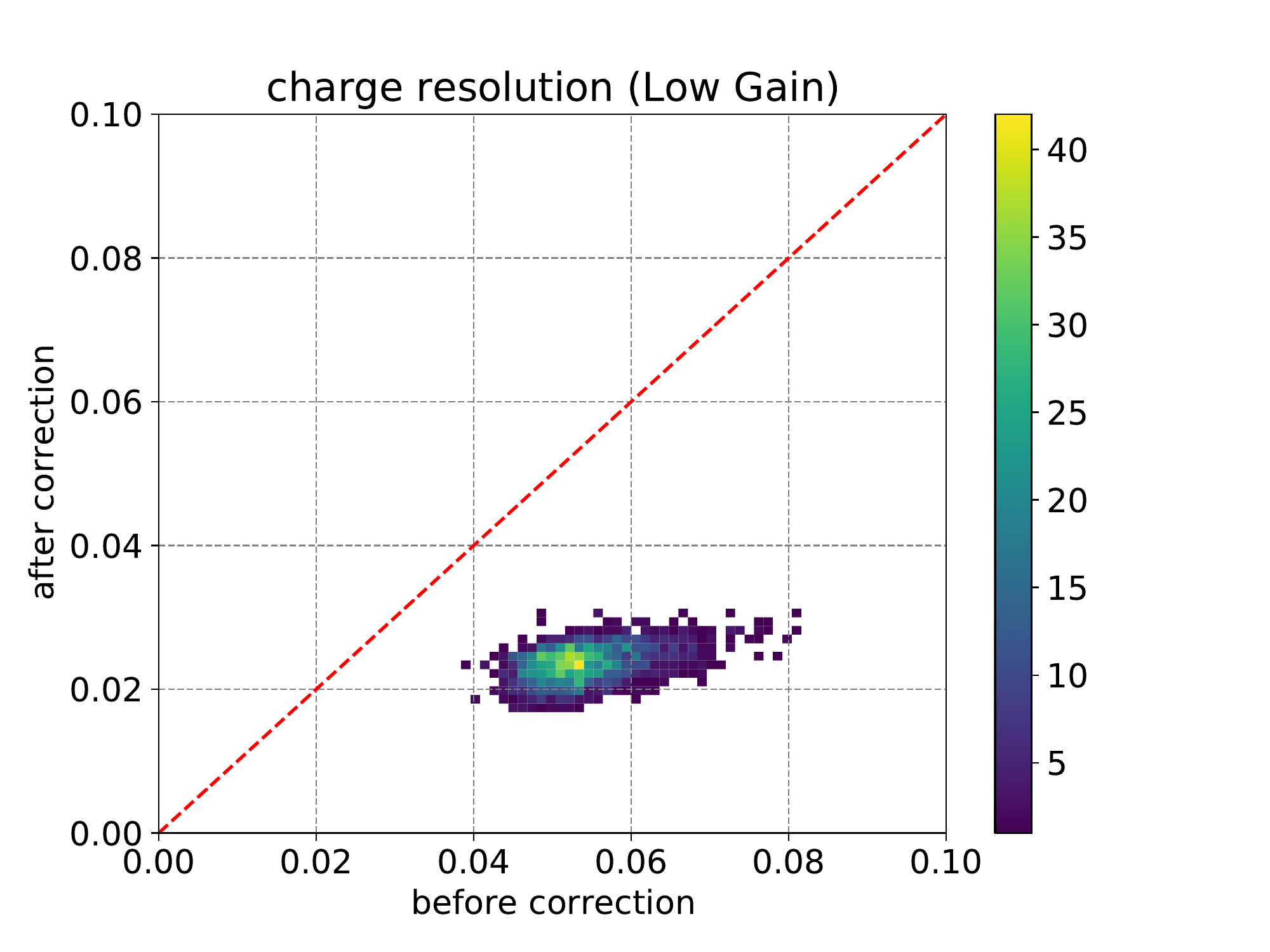}
  \end{minipage}
  \caption{Charge resolutions of all pixels measured with test pulses at high gain channels (left) and low gain channels (right) before and after the sampling interval correction.}
  \label{fig:charge_reso}
\end{figure}

Figure \ref{fig:charge_reso} shows the charge resolution of all pixels at both high-gain and low-gain channels with test pulses before and after the timing correction described in Section \ref{sec:time_correction_peak}. We used test pulses with an amplitude of 60--70 p.e. for high-gain channels and $\sim$1000 p.e. for low-gain channels. The charge resolutions at the high-gain and low-gain channels are 6--8\% and 4--6\% before the correction and 3--5\% and 2--3\% after the correction, respectively. This sampling interval correction improved the charge resolutions of all pixels.

\subsection{Time Resolution}
\begin{figure}
  \begin{minipage}[b]{0.5\linewidth}
    \centering
    \includegraphics[clip, height=6.0cm]{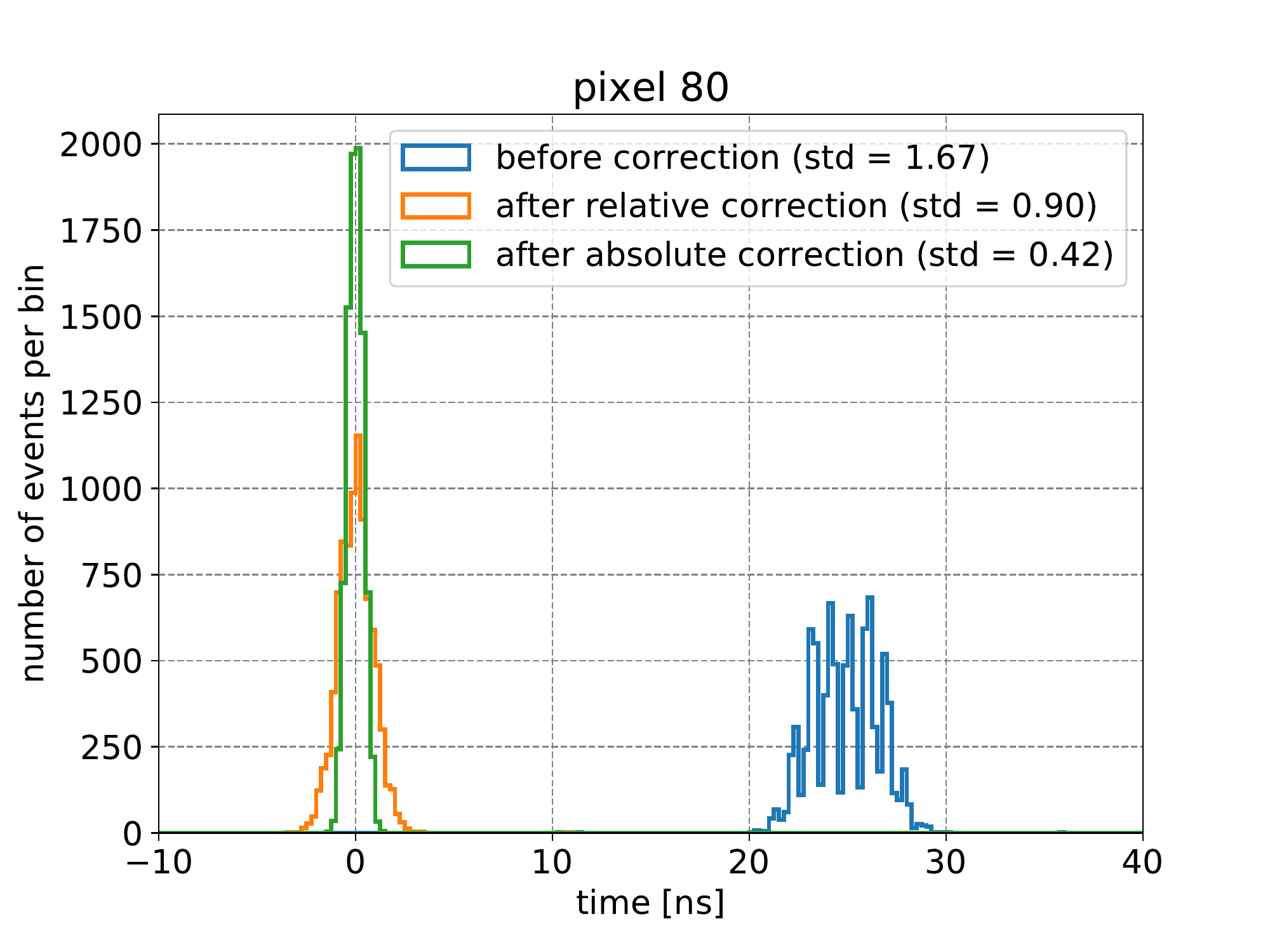}
  \end{minipage}
  \begin{minipage}[b]{0.5\linewidth}
    \centering
    \includegraphics[clip, height=6.0cm]{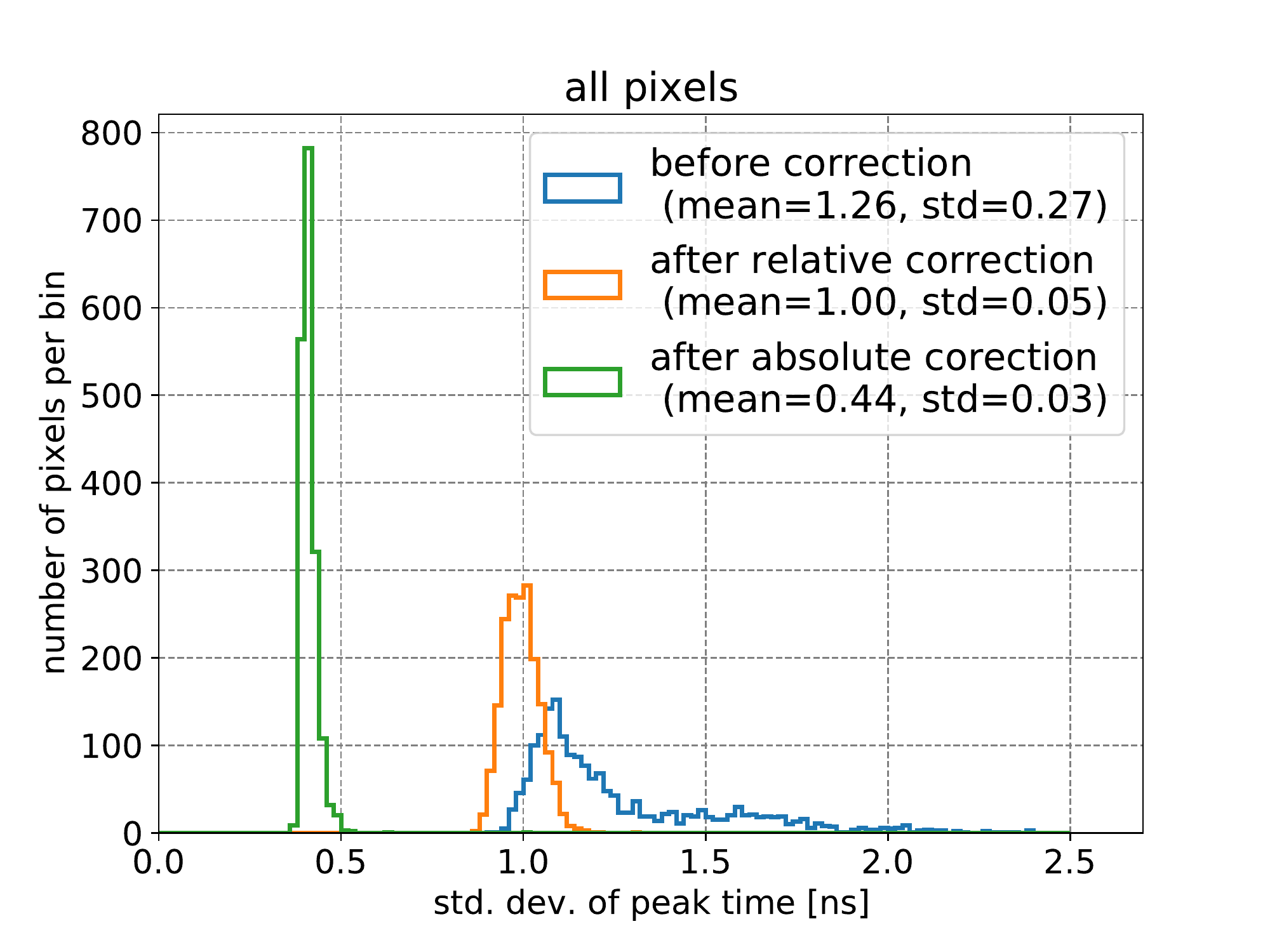}
  \end{minipage}
  \caption{Left: Distribution of pulse peak time with calibration laser at a given pixel before correction (blue), after relative correction described in Section \ref{sec:time_correction_peak} (orange), after absolute correction to compensate for the timing jitter of the trigger (green). Right: Distribution of a standard deviation of pulse peak time distribution of each pixel in the camera before correction (blue), after relative correction (orange), and after absolute correction (green).}
  \label{fig:time_reso}
\end{figure}

Figure \ref{fig:time_reso} (left) shows a distribution of pulse peak time with a calibration laser at a given pixel before and after peak time corrections. The pulse peak is given by the center of gravity between eight cells around the peak. Before the correction, the standard deviation is 1.7~ns. After the relative correction described in Section \ref{sec:time_correction_peak}, the standard deviation is improved to 0.9~ns. Absolute correction is done to compensate for the effect of timing jitter of the trigger or the calibration laser itself by subtracting the mean peak time among all pixels in the camera for every event. Figure \ref{fig:time_reso} (right) shows a distribution of the standard deviation of pulse peak distributions for all pixels in the camera after each step of the correction. Before the correction, the distribution has a tail component with poor time resolution. Those components disappeared after the correction. The mean value of this distribution referred to as time resolution is 1.0~ns after the relative correction and 0.4~ns after the absolute correction. 

\section{Summary}
\label{sec:summary}
We have developed the focal plane camera, which has 1855 PMTs and readout boards based on the ultra-fast waveform sampling DRS4 chips. The first LST was inaugurated in October 2018 in Spain, and various hardware calibration tests were performed. In parallel, we have been developing the analysis chain to compensate for the intrinsic features of the DRS4 chips. For this low-level calibration, it is firstly needed to extract parameters on the pedestal values for the time-lapse correction and cell-wise pedestal subtraction from the dedicated pedestal run with the random trigger. Furthermore, spike correction is also needed to remove the consecutive two spike cells. Sampling time inhomogeneity makes an effect on charge integration and pulse peak extraction. Thus, it is needed to prepare weighting factor and Fourier expansion parameters, respectively. 

After the dedicated low-level calibration, we evaluated the readout performance of the camera. The noise level of the readout board corresponds to 0.21 p.e. with 8-cell integration. Mean charge distribution taken with a calibration laser has a standard deviation of 1.9\% after the high-voltage flat-fielding. We confirmed that the sampling inhomogeneity correction is working well, and the charge resolution with test pulses is around 2--4\% after the correction. After the peak time correction to compensate for the chip-dependent features, the mean time resolution in the camera is 1.0~ns after the relative correction and 0.4~ns after the absolute correction.

\acknowledgments 
We gratefully acknowledge support from the agencies and organizations listed under Funding
Agencies at this website: http://www.cta-observatory.org/. We also acknowledge the great help from Open-It Consortium on the hardware development of the DRS4 readout board. SN is supported by JSPS KAKENHI Grant Number JP19J11940, and PG and JS are supported by the Narodowe Centrum Nauki grant No.2019/34/E/ST9/00224.

\bibliography{report} 
\bibliographystyle{spiebib} 

\end{document}